\documentclass[aps,prm,twocolumn,floatfix,10pt]{revtex4}
\usepackage{graphicx}
\usepackage{amsmath}
\usepackage{color}
\usepackage{wrapfig}
\usepackage{latexsym}
\begin{document}

\title{Thermal conductivity of polymers: A simple matter where complexity matters}

\author{Debashish Mukherji}
\email[]{debashish.mukherji@ubc.ca}
\affiliation{Quantum Matter Institute, University of British Columbia, Vancouver V6T 1Z4, Canada}

%\date{\today}

\begin{abstract}
Thermal conductivity coefficient $\kappa$ measures the ability of a material to conduct a heat current.
In particular, $\kappa$ is an important property that often dictates the usefulness of a material over a wide range of environmental conditions. 
For example, while a low $\kappa$ is desirable for the thermoelectric applications, a large $\kappa$ is 
needed when a material is used under the high temperature conditions. These materials range from common crystals 
to commodity amorphous polymers. The latter is of particular importance because of their use
in designing light weight high performance functional materials. In this context, however, one of the major limitations of the amorphous polymers is 
their low $\kappa$, reaching a maximum value of about 0.4 W/Km that is 2--3 orders of magnitude smaller 
than the standard crystals. Moreover, when energy is predominantly transferred through the bonded connections, $\kappa \ge 100$ W/Km. 
Recently, extensive efforts have been devoted to attain a tunability in $\kappa$ via macromolecular engineering. 
In this work, an overview of the recent results on the $\kappa$ behavior in polymers and polymeric solids is presented. 
In particular, computational and theoretical results are discussed within the context of complimentary experiments. 
Future directions are also highlighted. 
\end{abstract}

\maketitle

\section{Introductory remarks}
\label{sec:intro}

When a material is subjected to a non--equilibrium condition of temperature $T$, such that one end of the 
material is kept at an elevated temperature $T_{\rm high}$ and a lower temperature $T_{\rm low}$ is maintained at 
the other end, a heat current flows from the hot to the cold region. This is a direct consequence of the second law of thermodynamics and 
is quantified in terms of the heat flux vector $\vec{j}$. Here, the Fourier's law of heat diffusion
states that ${j} \propto \left( T_{\rm high} - T_{\rm low}\right)/\ell$ across a sample of length $\ell$.
The proportionality constant is thermal transport coefficient $\kappa$, which is a key material property that commonly 
dictates the usefulness of a material for a particular application~\cite{CHOY1977984,kappaGen2,PolRevTT14,kappaGen1,Kappabook,Keblinski20,Chen21Rev,kappaGen3}. 

Traditionally, heat flow in crystalline materials and in nano--structures have been of primary 
interest~~\cite{Olson93Sc,CahillRev03JAP,Boukai2008,Ren11NL,Mikolajick2013,Braun18AM}.
Due to the long range order in crystals, the phonon mean free path $\Lambda$ are rather large and thus $\kappa \ge 100$ W/Km~\cite{CrystalEXP,Si1,Fugallo18PS,Chen21Rev}. 
In the carbon--based materials, $\kappa$ can even exceed 1000 K/Wm~\cite{CNT1,CNT2,CNT3,CNT4}.
A complete opposite class to the crystals is the amorphous solids, where $\Lambda$ is  small, 
i.e, within the direct atom--to--atom contact. Here, $\kappa \le 2$ W/Km and heat propagates via localized vibrations~\cite{AmorphSi,keb09jap,pmmalocalized}. 

Within the class of amorphous solids, polymers are of particular importance because they usually provide a flexible platform for 
the design of advanced functional materials~\cite{Mueller20PPS,weil20rev,Mukherji20AR,nancy22rev}.
Some examples include, but are not limited to, organic solar cells~\cite{Ren11NL,Smith16AAMI,chen18ScAdv}, electronic packaging and/or heat sinking materials~\cite{Pipe15NMat,cahill16Mac,Smith16AAMI,AShank17ScAdv}, 
thermal switches~\cite{pnipamLCSTkappa,tian18acsml,thermS19jpcc,CisTrans19PNAS,ishida22NL} and thermoelectric applications~\cite{Shi17AFM,TherElecpol}.
However, the typical $\kappa$ values of the amorphous polymeric solids are further 5--10 times smaller~\cite{Pipe15NMat,cahill16Mac,Keblinski20,DMKKpolRev23} 
than the standard amorphous materials (such as amorphous silicon). This often hinders the usefulness of polymers under the high $T$ conditions.

\begin{figure}[ptb]
\includegraphics[width=0.46\textwidth]{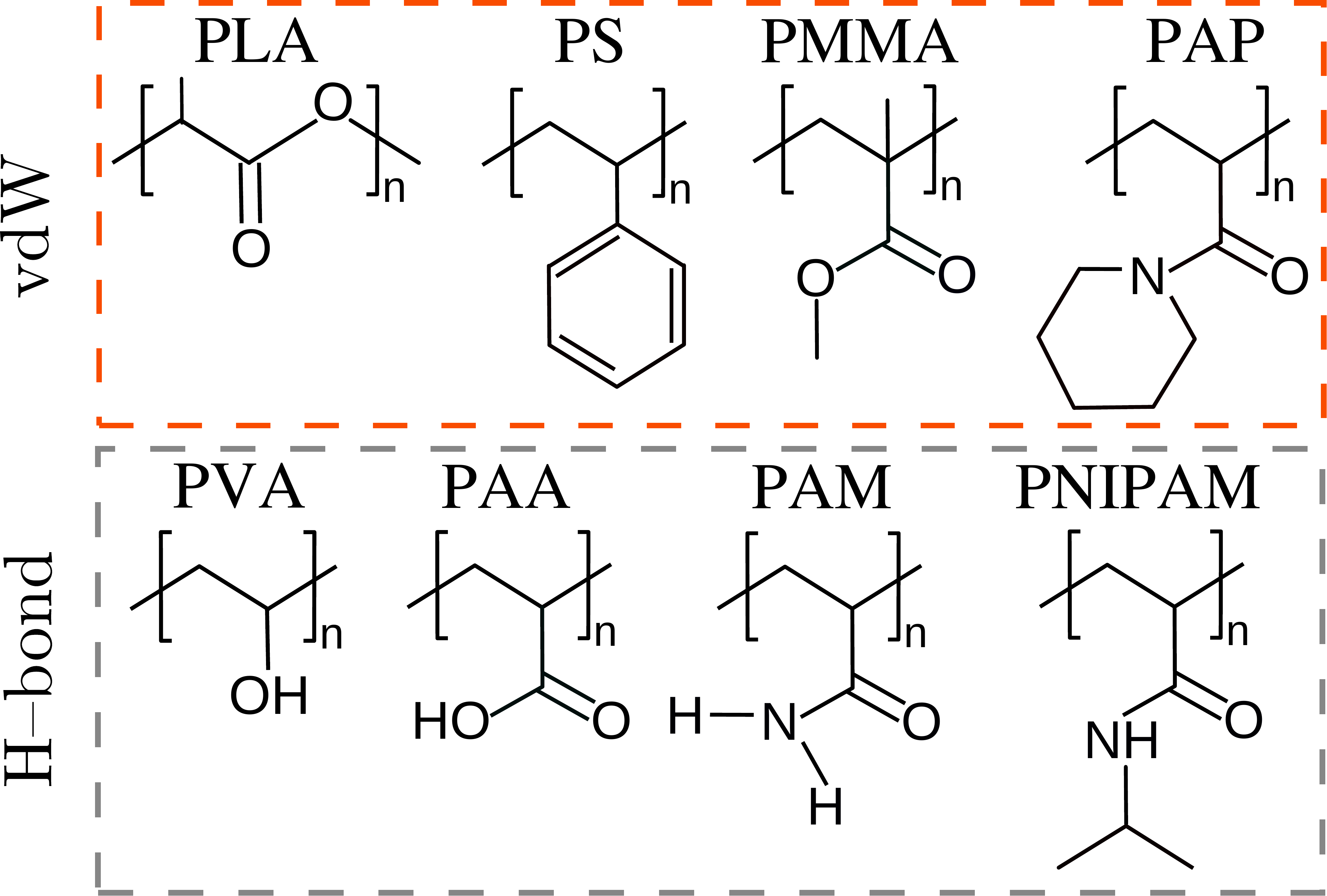}
\caption{Schematics of a few examples of different commodity polymeric structures. Top panel shows systems with
van der Waals (vdW) interactions, i.e., poly(lactic acid) (PLA), polystyrene (PS), poly(methyl methacrylate)(PMMA), and poly(N-acryloyl piperidine) (PAP).
Bottom panel shows hydrogen bonded (H--bond) systems, i.e., poly(vinyl alcohol) (PVA), 
poly(acrylic acid) (PAA), poly(acrylamide) (PAM), and poly(N-isopropyl acrylamide) (PNIPAM).}
\label{fig:struct}
\end{figure}

Most commonly known (non--conducting) commodity polymers can be categorized into the systems where non--bonded monomer--monomer 
interactions are either dictated by the van der Waals (vdW) forces or by the hydrogen bonds (H--bonds)~\cite{PH}. 
Here, the interaction strength of vdW is about $k_{\rm B}T$ at a temperature $T = 300$ K and the Boltzmann constant $k_{\rm B}$, 
while the strength of a H--bond is between 4--8$k_{\rm B}T$ depending on the dielectric constant of the medium~\cite{Mukherji20AR,desiraju02}.  
A few examples of the commodity polymers is shown in Fig.~\ref{fig:struct}.
Note that these particular polymers are chosen because their experimental and simulation data is readily available.

Experiments have reported that $\kappa \simeq 0.1-0.2$ W/Km for the vdW systems~\cite{Keblinski20,Pipe15NMat,pmma14exp,plakappa,PMMACpexp,pmmalocalized} 
and for the H--bonded polymers $\kappa \to 0.4$ W/Km~\cite{Pipe15NMat,cahill16Mac,Mukherji19PRM}.
\begin{table}[ptb]
        \caption{Thermal transport coefficient $\kappa$ for different commodity polymers and their
        corresponding glass transition $T_{\rm g}$. The data is compiled from the experimental
        literature, except for PNIPAM which is taken from simulation.
}
\begin{center}
       \begin{tabular}{|c|c|c|c|c|c|c|c|c|c|c|c|}
\hline
Interaction & Polymer         &   ~~~~$\kappa$  [W/Km]~~~~~    &    ~~~~$T_{\rm g}$  [K]~~~~  \\\hline
\hline
  vdW            & PLA              &  0.064--0.090 \cite{plakappa}    & 335 \cite{PH}   \\
              & PS              &   0.175 \cite{keb09jap}   &  373 \cite{PH}   \\
              & PMMA            &   0.200 \cite{cahill16Mac}   &  378 \cite{PH}    \\
              & PAP             &   0.160~\cite{cahill16Mac} \& 0.200~\cite{Pipe15NMat}    &  380 \cite{Pipe15NMat}   \\
               \hline
               \hline
               \hline
  H--bond            & PVA             &   0.310 \cite{cahill16Mac}   &  348  \cite{PH}  \\
              & PAA             &   0.370 \cite{cahill16Mac}   &  385  \cite{Pipe15NMat}  \\
              & PAM             &   0.380 \cite{cahill16Mac}   &  430  \cite{PH}  \\
              & PNIPAM          &   $0.316^{\rm sim}$ \cite{Mukherji19PRM}    &  413  \cite{Mukherji19PRM}  \\
\hline
\end{tabular}  \label{tab:tg_kappa}
\end{center}
\end{table}
Table~\ref{tab:tg_kappa} lists $\kappa$ and the corresponding glass transition temperatures $T_{\rm g}$ for the polymers in Fig.~\ref{fig:struct}. 
It can be seen in Table~\ref{tab:tg_kappa} that PMMA (a vdW--based polymer) has $\kappa \simeq 0.20$ W/Km 
and its $T_{\rm g} \simeq 378$ K, while PVA (a H--bonded system) has $\kappa \simeq 0.310$ W/Km 
and a lower $T_{\rm g} \simeq 348$ K~\cite{PH}. These values indicate that $T_{\rm g}$ and $\kappa$ are not correlated, which is also visible across many polymeric systems~\cite{PH}. Furthermore, a closer look suggests that-- within a simple approximation, $T_{\rm g}$ is directly related to the trans--to--gauche free energy barrier $\Delta E_{\rm t-g}$ (i.e., local fluctuations),
which is dictated by a delicate combination of the bonded, the angular and the dihedral interactions along a chain backbone. The higher the $\Delta E_{\rm t-g}$, the larger the $T_{\rm g}$. Here, PMMA with a larger side group has a higher $\Delta E_{\rm t-g}$ than a PVA, hence the observed trend in $T_{\rm g}$. Within this simple discussion it becomes reasonably apparent that the exact $T_{\rm g}$ is a completely irrelevant quantity within the context of $\kappa$ in polymers.

\begin{figure}[ptb]
\includegraphics[width=0.49\textwidth]{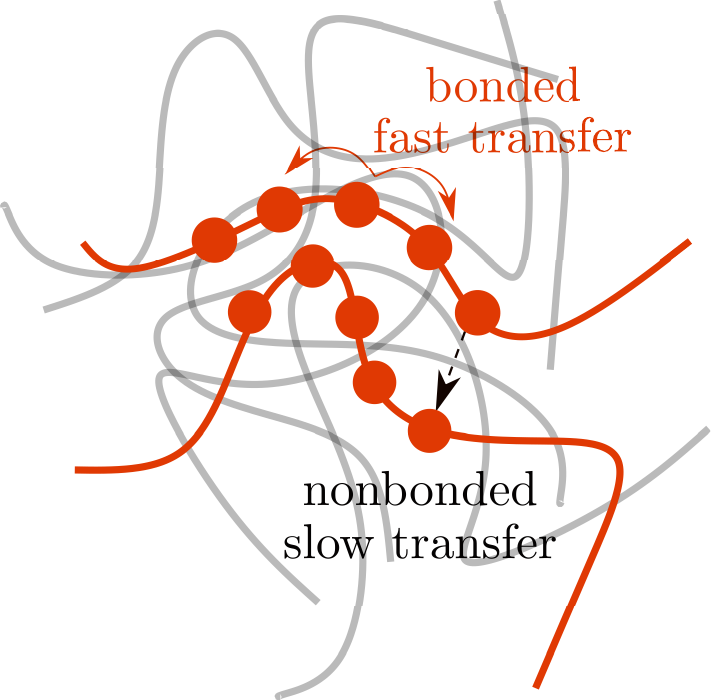}
\caption{A schematic representation of the chain configurations in a melt. A fast energy transfer rate between 
two bonded and slow energy transfer between two non--bonded monomers are shown by the orange and black arrows, respectively.}
\label{fig:schemCETM}
\end{figure}

Amorphous polymeric solids are a special case because even when their macroscopic $\kappa$ values are very small, 
at the monomer level they have different rates of energy transfer. For example, energy can be transferred 
between the bonded monomers and that between the neighboring non--bonded monomers. 
In this context, a closer investigation of the polymer structures reveal that the carbon--carbon (C--C) covalent bond 
constitute the most common backbone of commodity polymers, see Fig.~\ref{fig:struct}. 
Here, it is known that the stiffness of a C--C contact is $E \ge 250$ GPa~\cite{CCBondE}, while $E$ of a vdW or a H--bond 
system vary between 2--5 GPa~\cite{cahill16Mac}. Given that $\kappa \propto E$~\cite{Cahill90PRB,Braun18AM}, the energy transfer between 
the two non--bonded monomers (soft contacts) is significantly smaller than along a (stiff) bond~\cite{MM21mac}. A simple schematic of this scheme is shown in Fig.~\ref{fig:schemCETM}.
%In an amorphous polymer, e
The thermal behavior in polymers are predominantly dictated by the non--bonded contacts, while the energy transfer along a bonded contact plays a lesser important role. This is particularly because a chain in a frozen configuration follows the random walk statistics~\cite{DGbook,DoiBook,DesBook}, i.e., when it is quenched to $T \ll T_{\rm g}$ 
from a melt configuration.
Within this picture, when heat flows along a chain contour, it experiences scattering due to the bends and the kinks 
along the path~\cite{chen19jap,bhardwaj2021thermal}. Energy also occasionally hops off to a non--bonded neighboring monomer. 
A combination of these two effects facilitates a knocking down of $\kappa$ in the amorphous polymers~\cite{keb09jap,pmmalocalized,MukherjiarXiv}. Note also that the exact monomer structure, i.e., the side groups connected to a backbone play an additional (delicate) role~\cite{kappaHaoMaa,kappaside18mat,mukherji24lang}, which will be discussed at a later stage within this short overview.

Over the last 2--3 decades, extensive efforts have been devoted to study the heat flow in polymers using the experimental, theoretical, and computational approaches~\cite{shen2010polyethylene,Pipe15NMat,cahill16Mac,Smith16AAMI,tomko2018tunable,tian18acsml,Bruns19mac,thermS19jpcc,PolRev20JHT,DMPRM21,Cahill21SemiCrys,mukherji22acsml}. 
In particular, even when the polymers are a class of simple matter with their great potential in designing flexible materials, establishing a microscopic understanding in polymers is rather complex. Here, one of the grand challenges in this field is to attain a predictive tunability in $\kappa$ (almost {\it at will}) using macromolecular engineering.
This requires a protocol that can properly account for a delicate balance between the bonded to the non--bonded interactions, the chain conformations, 
and their morphology. Motivated by the above, this manuscript aims to highlight the latest developments in the field of polymer thermal conductivity. 
For this purpose, comparative experimental and simulation results will be discussed to put forward the key concepts.

\section{Effect of blending on the thermal conductivity of polymers}

In a system consisting of one (linear) polymer component, $\kappa$ is rather limited because of the 
restrictive monomer--level interactions~\cite{cahill16Mac,Pipe15NMat,Mukherji19PRM}. To circumvent this problem, 
studies have suggested that $\kappa$ of a polymeric solid may be enhanced by blending a second component with a relatively higher $\kappa$. 
Here, an obvious choice is the carbon--based materials, such as the carbon nanotube (CNT) and polymer composites~\cite{Smith16AAMI,Winney06mac,CNTPolComp11Rev}. 
In such a composite, a significant increase in $\kappa$ requires concentration of CNT $\phi_{\rm CNT}$
exceeding their typical percolation threshold. While a CNT--polymer composite certainly show a significantly higher 
$\kappa$ than the bare polymers~\cite{Smith16AAMI,CNTPolComp11Rev}, it also has two major drawbacks: (1) it looses the underlying flexibility (typical of polymers) because of a large $\phi_{\rm CNT}$ and their physical properties 
then get dominated by the CNTs present in the background polymers. 
(2) Polymers are rather cost effective, having their typical 
prices of about 2--3 orders of magnitude lower than the CNTs
and thus a CNT--polymer composite inherently becomes significantly costlier. 

A more plausible alternative is the polymer blends,
where the non--bonded interactions can be altered by changing $\phi_{\rm second}$ of the second polymer component.
Here, however, a prerequisite is that the two components remain fairly miscible over the full range of $\phi_{\rm second}$~\cite{cahill16Mac,Pipe15NMat,he21acsmac,Jayaraman22mac,mukherji22prm}. 
On the contrary, when the two components in a blend phase separate, they create zones within a sample 
consisting of the individual components. These separate zones usually have very weak interfacial interaction 
and thus induce resistance for the heat flow, akin of the Kapitza resistance~\cite{KapitzaResRev}.

In a symmetric polymer blends (i.e., when both polymer components have a comparable degree of polymerization $N_{\ell}$), 
$\kappa$ varies monotonically between the two pure components~\cite{cahill16Mac,mukherji22prm}. 
A recent experimental study, however, reported that an asymmetric blend (consisting of the longer PAA and shorter PAP chains) shows a larger enhancement in $\kappa \simeq 1.5$ W/Km around a PAP concentration of $\phi_{\rm PAP} \simeq 30\%$~\cite{Pipe15NMat}. 
It was argued that the PAP chains act as the H--bonded cross--linkers between the neighboring PAA chains, 
forming a 3--dimensional H--bonded stiff network.
On the contrary, another set of experiments did not attain the same enhancement, instead found that PAA--PAP phase separate around $\phi_{\rm PAP} \simeq 30\%$~\cite{cahill16Mac}.

Motivated by the above contradicting experimental results, a simulation study suggested 
that the miscibility can be enhanced when PAP is replaced with PAM, i.e., a PAA--PAM system consisting of 
a long PAA and a short PAM~\cite{Bruns19mac}. 
PAA--PAM showed a weak non--monotonic variation in $\kappa$ with $\phi_{\rm PAM}$, attaining a maximum $\kappa \simeq 0.4$ W/Km around $\phi_{\rm PAM} \simeq 30\%$. 
Note also that the size of PAM molecules in this simulation study was chosen to be of the order of persistence length $\ell_{\rm p} \simeq 0.75$ nm (or 3 monomers), 
i.e., a PAM as a stiff linker that fits perfectly between two PAA chains. 
When the cross--linker length $N_{\ell} \gg \ell_{\rm p}$, they form flexible cross--linking. This on one hand makes a network soft~\cite{lv2021effect}, on the other they also induce effective free volume (weak spots) within a network~\cite{DMPRM21}. Collectively, these two effects reduce $\kappa$. Something that speak in this favor is that the experimental results of cross--linked PAA reported $\kappa \simeq 0.27$ W/Km, which is about 25\% smaller than $\kappa \simeq 0.37$ W/Km measured in a linear PAA~\cite{cahill16Mac}. 

The above discussions suggest that there is a need to look beyond the simple amorphous polymers. Therefore, in the 
following section some analytical approaches are first presented that may provide a guiding tool for
the remaining discussions presented herein.

\section{Analytical models} %of thermal conductivity in non--conducting polymers}

In an isotropic material, $\kappa$ is directly related to the volumetric heat capacity $c$, 
the group velocity $v_{{\rm g},i}(\nu)$, and the phonon mean--free path $\Lambda(\nu) = \tau(\nu) v_{{\rm g},i}(\nu)$.
Here, $\tau(\nu)$ is the phonon life time and $\nu$ is the vibrational frequency. 
Starting from the above description, $\kappa$ can be written as,
\begin{equation}
    \kappa(\nu) = \frac {1}{3} \sum_i c(\nu) v_{{\rm g},i}^2(\nu) \tau(\nu).
    \label{eq:kappa_gen}
\end{equation} 
For an non--conducting amorphous material a well known theoretical 
description is the minimum thermal conductivity model (MTCM)~\cite{Cahill90PRB} that is discussed in the following.

\subsection{The minimum thermal conductivity model}
\label{ssec:mtcm}

Following Eq.~\ref{eq:kappa_gen}, the general expression of $\kappa$ for a 3--dimensional isotropic system reads~\cite{Kappabook},
\begin{equation}
    \kappa = \left (\frac {\rho_{\rm N} h^2} {3k_{\rm B} T^2}\right)
    \sum_i \int \tau(\nu) v_{{\rm g},i}^2(\nu) 
    \frac {\nu^2 e^{{h\nu}/{k_{\rm B}T}}} {\left(e^{{h\nu}/{k_{\rm B}T}} -1 \right)^2}
    g(\nu) {\rm d}\nu,
    \label{eq:k_3d}
\end{equation}
where $\rho_{\rm N}$, $g(\nu)$, $h$ are the total particle number density, the vibrational density of states, and the Planck constant, respectively. 
Within this description, MTCM uses the Debye model of lattice vibrations in Eq.~\ref{eq:k_3d} and proposes 
that a sample can be divided into regions of size $\Lambda(\nu)/2$, whose frequencies are given by the low $\nu$ sound wave velocities $v_{i}$, 
and thus approximates $\tau = 1/2\nu$~\cite{Cahill90PRB} and $v_{{\rm g},i} = v_i$. This gives,
\begin{equation}
    \kappa = \left (\frac {\rho_{\rm N} h^2} {6k_{\rm B} T^2}\right) %_{\rm MTCM}
    \left(v_{\ell}^2 + 2v_{t}^2 \right) \int {\mathcal I}(\nu)
    g(\nu) {\rm d}\nu,
    \label{eq:k_mtm}
\end{equation}
with
\begin{equation}
    {\mathcal I}(\nu) = \frac {\nu e^{{h\nu}/{k_{\rm B}T}}} {\left(e^{{h\nu}/{k_{\rm B}T}} -1 \right)^2}.
    \label{eq:k_mtm2}
\end{equation}
$v_{\ell} = \sqrt{C_{\rm 11}/\rho_{\rm m}}$ and $v_{t} = \sqrt{C_{\rm 44}/\rho_{\rm m}}$ are the longitudinal and the transverse sound wave velocities, respectively. 
Here, $C_{\rm 11} = K + 4 C_{\rm 44}/3$, $K$ is the bulk modulus, $C_{\rm 44}$ is the shear modulus, and $\rho_{\rm m}$ is the mass density.

\begin{figure}[ptb]
\includegraphics[width=0.49\textwidth]{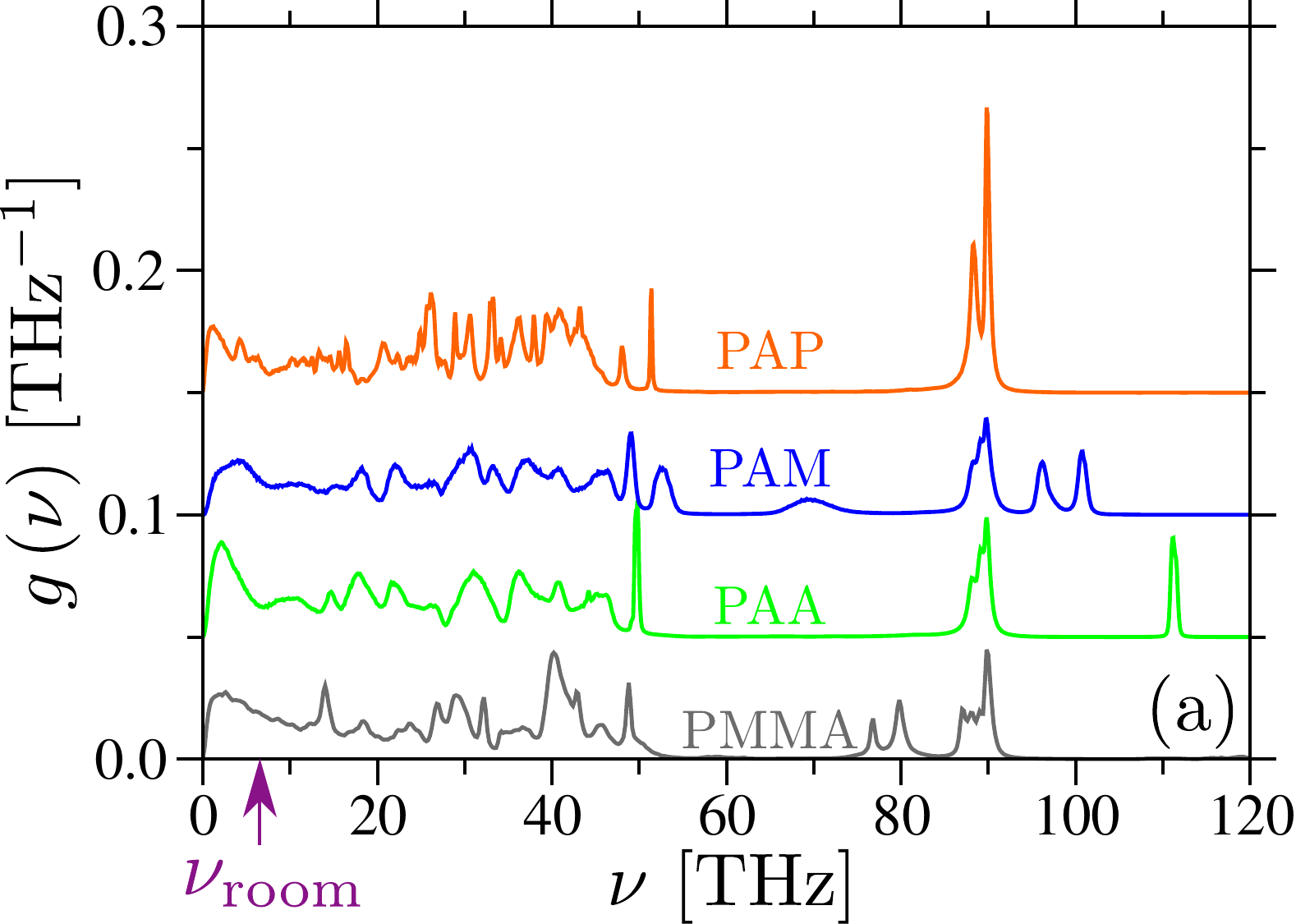}
\caption{Vibrational density of states $g(\nu)$ for four commodity polymers, 
namely; poly(N-acryloyl piperidine) (PAP), polyacrylamide (PAM), poly(acrylic acid) (PAA), and poly(methyl methacrylate)(PMMA).
Individual $g(\nu)$ are shifted for a clearer representation. 
The arrow indicates the characteristic frequency $\nu_{\rm room} \simeq 6.2$ THz at a temperature $T = 300$ K.
This figure is reproduced with permission from Ref.~\cite{MukherjiarXiv}.} 
\label{fig:gnu_4pol}
\end{figure}

One key quantity in Eq.~\ref{eq:k_mtm} is $g(\nu)$, which can be calculated by the Fourier transform
of the mass--weighted velocity auto--correlation function $c_{\rm vv}(t) = \sum_{i} m_i \langle {\overrightarrow v}_i(t) \cdot {\overrightarrow v}_i(0)\rangle$~\cite{Horbach1999JPCB,martin21prm}
obtained from the classical simulations,
\begin{equation}
    g(\nu) = \frac {1}{C}\int_{0}^{\infty} \cos(2\pi \nu t) \frac{c_{\rm vv}(t)}{c_{\rm vv}(0)} {\rm d}t.
    \label{eq:vdos}
\end{equation}
The prefactor $C$ ensures that $\int g(\nu){\rm d}\nu = 1$. The representative $g(\nu)$ for four different commodity polymers are shown
in Fig.~\ref{fig:gnu_4pol}.
It can be appreciated that the polymers have many high $\nu$ quantum degrees--of--freedom that do not contribute to $\kappa$ at $T = 300$ K. 
For example, a C--H bond vibration frequency in a polymer is $\nu \simeq 90$ THz, while the representative $\nu_{\rm room} \simeq 6.2$ THz at $T = 300$ K. 
Such a mode, together with many other stiff modes (for $\nu > \nu_{\rm room}$), remain quantum--mechanically frozen at $T = 300$ K~\cite{martin21prm}. 
If the contributions of these individual modes are not properly incorporated in Eq.~\ref{eq:k_mtm} via the Bose--Einstein weighted function in Eq.~\ref{eq:k_mtm2}, 
one can easily overestimate $\kappa$ within the classical simulations in comparison to the experimental data~\cite{Mukherji19PRM,KappaMDExp,martin21prm,kappaOil,MukherjiarXiv}.   

Standard analytical descriptions typically use the Debye estimate of parabolic vibrational density of states $g_{\rm D}(\nu) = 3 \nu^2/\nu_{\rm D}^3$
in Eq.~\ref{eq:k_mtm}. Here, $\nu_{\rm D}$ is the Debye frequency and is written as~\cite{Kappabook,Horbach1999JPCB},
\begin{equation}
    \nu_{\rm D} = \left( \frac {9 \rho_{\rm N}} {4\pi} \right)^{1/3} \left( \frac {1} {v_{\ell}^3} + \frac {2}{v_{t}^3} \right)^{-1/3}.
\end{equation}
While $g_{\rm D}(\nu)$ is certainly a good approximation for the standard amorphous solids when
$T \ll \Theta_{\Theta}$, with $\Theta_{\rm D} = h\nu_{\rm D}/k_{\rm B}$ being the Debye temperature. 
Typical examples are amorphous silica and/or silicon, where $\Theta_{\rm D} \ge 480$ K~\cite{Horbach1999JPCB,DebyeTaSi,DebyeTaSi1}.
Moreover, in the case $g(\nu)$ is complex, such as in the polymers (see Fig.~\ref{fig:gnu_4pol}), 
simplistic $g_{\rm D}(\nu)$ may lead to a wrong estimate of the low $\nu$ vibrational modes 
and thus leads to the default artifacts in computed $\kappa$. This is particularly because $\Theta_{\rm D} \simeq 180-220$ K 
for the commodity polymers listed in Table~\ref{tab:tg_kappa}, i.e., 20--40\% smaller than $T = 300$ K~\cite{cahill16Mac,MukherjiarXiv}
where typical experiments and simulations are performed.

\begin{figure}[ptb]
\includegraphics[width=0.49\textwidth]{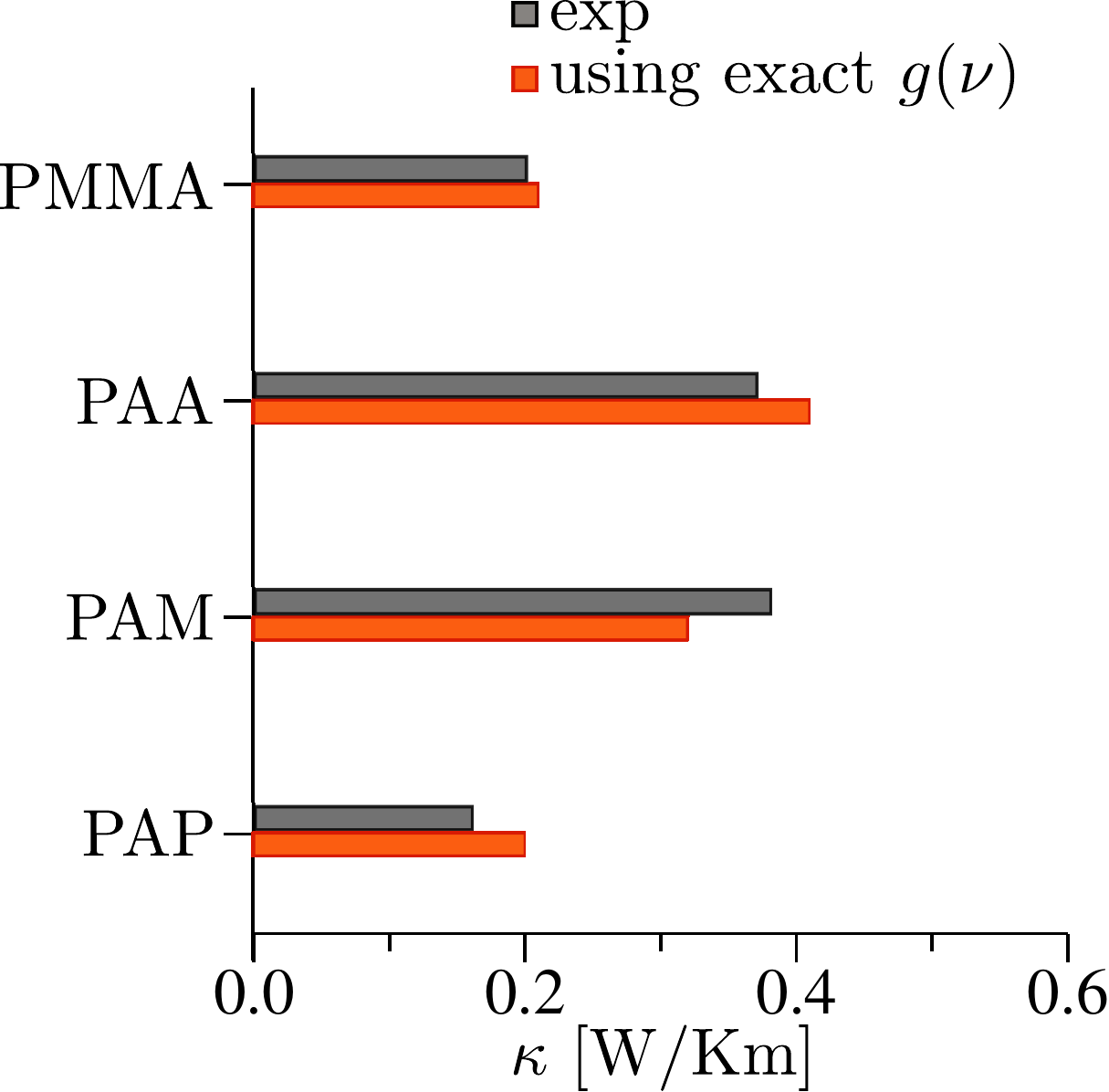}
\caption{A plot comparing the $\kappa$ values calculated using Eq.~\ref{eq:k_mtm} and the 
corresponding experimentally measured $\kappa$~\cite{cahill16Mac}. 
Data for four commodity polymers are shown, namely; poly(N-acryloyl piperidine) (PAP), polyacrylamide (PAM), poly(acrylic acid) (PAA), and poly(methyl methacrylate)(PMMA).
This figure is reproduced with permission from Ref.~\cite{MukherjiarXiv}.}
\label{fig:k_dmi_exp}
\end{figure}

When exact $g(\nu)$ from Fig.~\ref{fig:gnu_4pol} are used in Eq.~\ref{eq:k_mtm}, $\kappa$ values can be reasonably 
reproduced within 5--20\% error, see Fig.~\ref{fig:k_dmi_exp}.
Note that this data is obtained by taking $v_{\ell}$ and $v_t$ from the experiments~\cite{cahill16Mac}, 
while $\rho_{\rm N}$ and $g(\nu)$ are taken from simulations~\cite{MukherjiarXiv}.

\subsubsection{High temperature approximation with correction of the stiff modes}

Within the high $T$ classical limit, i.e., when all modes are considered in Eq.~\ref{eq:k_mtm}~\cite{Cahill11PRB},
the original MTCM for amorphous polymers can be written as,
\begin{align}    \label{eq:k_mtmPol}
    \kappa_{\rm MTCM} &= \left (\frac {\pi} {48} \right)^{1/3} k_{\rm B} {N}^{2/3} %
    \left (v_{\ell} + 2v_{t} \right). %\\\nonumber
    %                   &= \left (\frac {\pi} {432} \right)^{1/3} k_{\rm B}^{1/3} {c}^{2/3}
    % \left (v_{\ell} + 2v_{t} \right).
\end{align}
The corrections for the stiff modes in a polymer can then be incorporated in Eq.~\ref{eq:k_mtmPol} by considering an effective number of 
atoms ${\overline N} = 2(N - N_{\rm H})/3$ and $c = 3 {\overline N} k_{\rm B}$. Here, ${\overline N}$ eliminates the stiff modes associated 
with the number of hydrogen atoms $N_{\rm H}$ and other stiff backbone modes. Following this, Eq~\ref{eq:k_mtmPol} can be simply written as,
\begin{align}    \label{eq:k_mtmPol2}
    \kappa_{\rm MTCM} = \left (\frac {\pi} {432} \right)^{1/3} k_{\rm B}^{1/3} {c}^{2/3}
    \left (v_{\ell} + 2v_{t} \right).
\end{align}
Eq.~\ref{eq:k_mtmPol2} gives estimates that are 30\% larger than the corresponding experimental data~\cite{cahill16Mac,Mukherji19PRM,lv2021effect}.

\subsubsection{Accurate computation of $\kappa$ via specific heat correction}

\begin{figure}[ptb]
\includegraphics[width=0.46\textwidth]{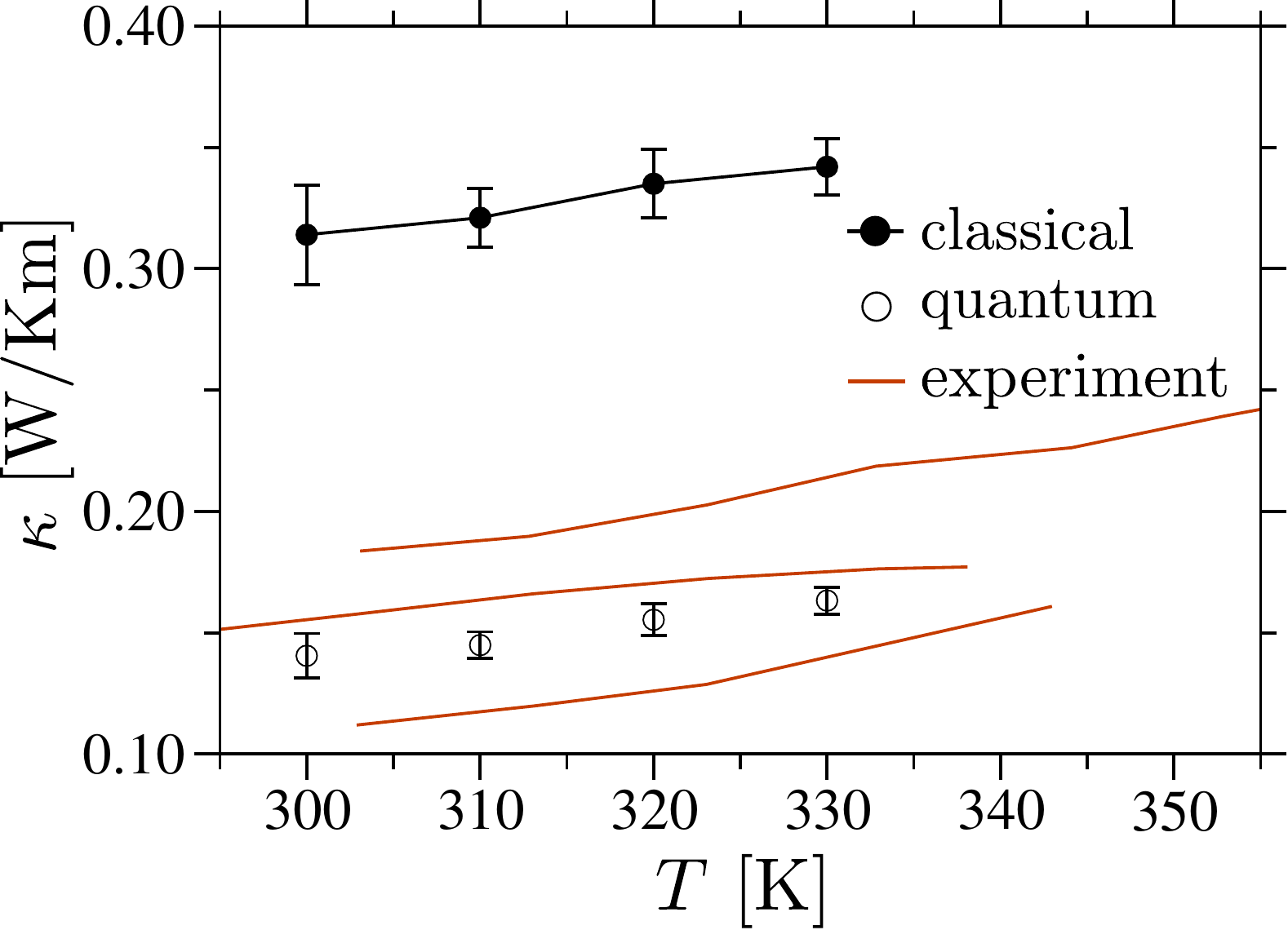}
\caption{Thermal transport coefficient $\kappa$ as a function of temprature $T$ for poly(methyl methacrylate) (PMMA).
The data sets are shown that correspond to the classically computed $\kappa_{\rm cl}$, quantum corrected $\kappa$ using
Eq.~\ref{eq:kt_cl} and the corresponding experiments.
This figure is reproduced with permission from Ref.~\cite{MukherjiarXiv}.}
\label{fig:kappaT}
\end{figure}

% The very simple microscopic detail discussed in the preceding paragraph also leads to a fundamental 
% artifact when $\kappa$ is calculated in the 
%using any common non--equilibrium method, such as the approach--to--equilibrium.

An accurate computation of $\kappa$ within the standard classical molecular simulation setups is a daunting
task, because polymers have quantum degrees--of--freedom whose exact contribution to the heat balance is rather non--trivial.
Furthermore, $T > \Theta_{\rm D}$ for the commodity polymers (see Section~\ref{ssec:mtcm})
and thus the polymer thermal properties are dominated by the low $\nu$ classical modes that are 
dominated by the non--bonded interactions (or the localized vibrations). On the contrary, the stiff modes (i.e., for $\nu > \nu_{\rm room}$) remain
quantum--mechanically frozen and do not contribute to $\kappa$. In this context, one of the key quantities 
that dictates $\kappa$ behavior in Eq.~\ref{eq:kappa_gen} for polymer is its $c$.

In classical simulations, every mode in a polymer contribute equally to $c$, i.e., given by the Dulong--Petit classical estimate.
Here, it is well documented that the classically computed $c$ are always overestimated in comparison to the corresponding 
experimental data~\cite{Horbach1999JPCB,BHOWMIK2019176,martin21prm,baschnagel22jcp} and thus
also leads to an overestimation of $\kappa$~\cite{KappaMDExp,kappaOil,MukherjiarXiv}.

Given the above discussion, if $c$ is estimated accurately, it will automatically lead to an accurate computation of $\kappa$.
Recently a method has been proposed to compute the quantum corrected $c$. This method uses the Binder approach to estimate the 
contributions of the stiff harmonic modes~\cite{Horbach1999JPCB}, which is then used to get the difference between the classical
and the quantum descriptions~\cite{martin21prm},
\begin{equation} \label{eq:shrel}
    \frac {\Delta c_\textrm{rel}(T)} {k_{\rm B}} 
    = \int_0^\infty \left\{ 1 - 
    \left(\frac {h\nu}{k_\textrm{B}T}\right)^2 \frac {e^{{h\nu}/{k_{\rm B}T}}} {\left(e^{{h\nu}/{k_{\rm B}T}} -1 \right)^2} \right\} g(\nu) {\rm d}{\nu}.
\end{equation}
Finally the quantum corrected estimate of $c(T)$ is can be given by,
\begin{equation}\label{eq:ccor}
    c(T) = c_{\rm cl}(T) - \Delta c_\textrm{rel}(T).
\end{equation}
Here, the classical heat capacity is calculated using $c_{\rm cl} = [H(T- \Delta T) - H(T - \Delta T)]/2 \Delta T$ 
and $H(T)$ is enthalpy. The main advantage of using Eq.~\ref{eq:ccor} is that the stiff harmonic modes 
are corrected, while the contributions from the anharmonic (low $\nu$) modes remain unaffected.

Using Eq.~\ref{eq:ccor}, one can then calculated the quantum corrected $\kappa(T)$,
\begin{equation}
\label{eq:kt_cl}
    \kappa(T) = c(T)\frac {\kappa_{\rm cl}(T)} {c_{\rm cl}(T)}.
\end{equation}
Here, $\kappa_{\rm cl}(T)$ is classically computed thermal transport coefficient using the standard equilibrium~\cite{ZwanzigRev} and/or non--equilibrium~\cite{MPkappa,Lampin2013} methods.
Fig.~\ref{fig:kappaT} show the computed $\kappa(T)$ for PMMA using $c(T)$~\cite{MukherjiarXiv}.
It can be seen that the quantum corrected $\kappa(T)$ compares reasonably with the corresponding experimental data, 
while the classical estimate is about a factor of three too high.
The method proposed in Ref.~\cite{martin21prm} also highlighted different strategies to estimate $c(T)$ 
accounting for the missing degrees--of--freedom (DOF) within the united--atom and/or coarse--grained models. 
A direct implication is that a certain percentage error in $\kappa$ computed in the united--atom models comes from
the missing DOFs~\cite{wu22cms}.

It is also important to highlight that the simple scaling in Eq.~\ref{eq:kt_cl} works reasonably for polymers because 
only the low $\nu$ modes dominate their thermal properties. When dealing with the crystalline materials special attention
is need. For example, in a crystal, not only $c$ that has quantum effects, rather $v_{\rm g}$ (i.e., stiffness)~\cite{MHM01prb} and $\tau$ 
also has quantum contributions at low $T \ll \Theta_{\rm D}$.

\subsection{Single chain energy transfer}
\label{ssec:cet}

In the introduction, it is discussed that a C--C bond is significantly stiffer~\cite{CCBondE} than the typical non--bonded contacts.
A direct consequence of this microscopic interaction contrast is that the energy transfer between two bonded monomers is over 100 times faster than 
the energy transfer between two non--bonded monomers~\cite{MM21acsn,MM21mac}. Taking motivation from such distinct microscopic interactions, 
experimental and computational/theoretical studies have reported a large enhancement in $\kappa$ in the systems where bonded 
interactions dominate, such as in the single extended chains~\cite{chen19jap,SingleExtend,PEDOT18prm,mukherji24lang}, polymer fibers~\cite{shen2010polyethylene,luo13acsnano}, 
and/or molecular forests~\cite{bhardwaj2021thermal,ThermConductScience}. However, until recently there existed no direct theoretical 
framework that could quantitatively decouple the effects of these two separate microscopic interactions in dictating the macroscopic heat flow in polymers~\cite{MM21acsn,MM21mac}.
Therefore, in this section, the key ingredients of this simple chain energy transfer model (CETM) will be discussed.

\begin{figure}[ptb]
\includegraphics[width=0.49\textwidth]{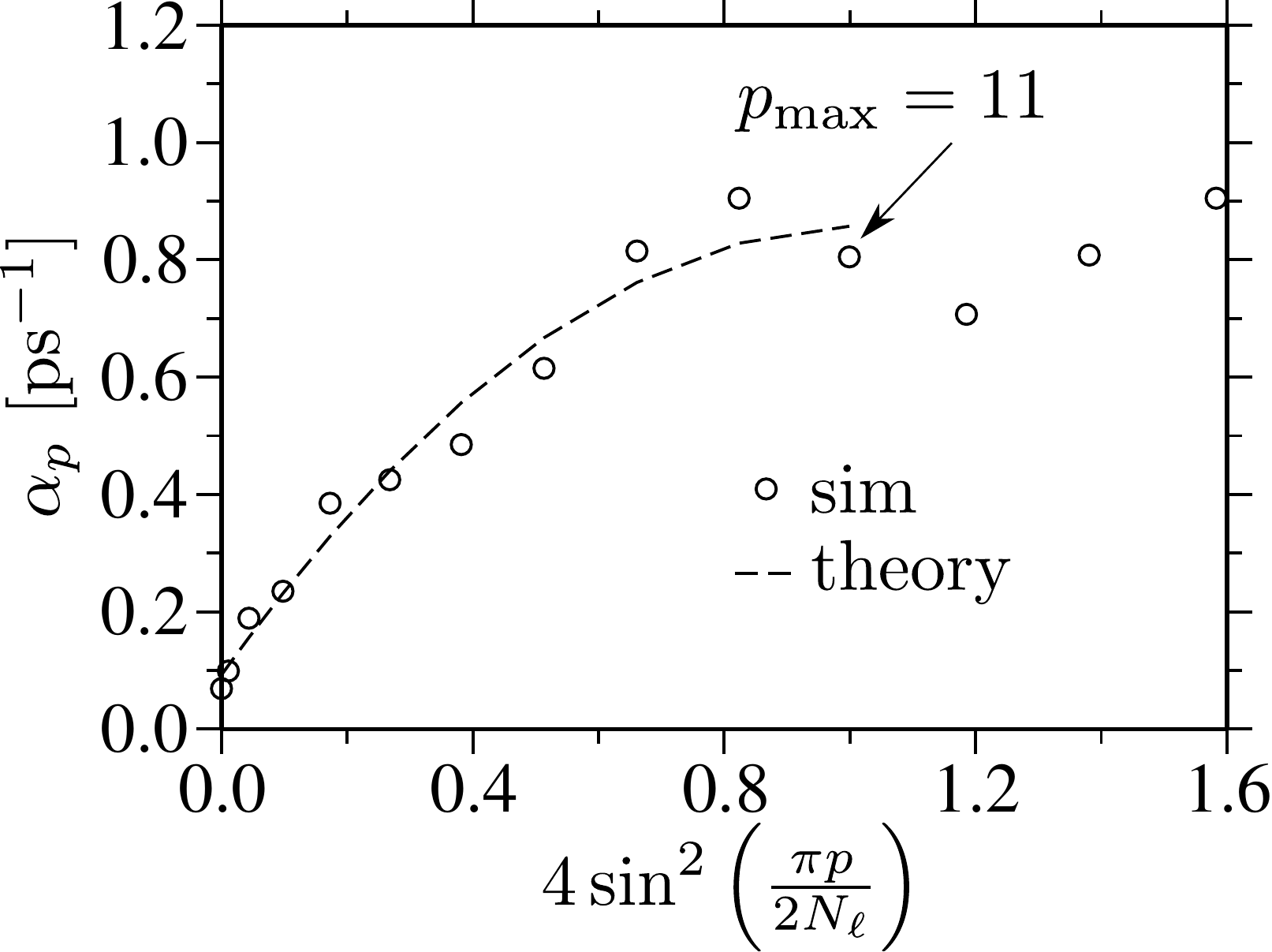}
\caption{Relaxation time constant $\alpha_p$ of the individual $p$ modes. In the $x-$axis, $p$ values are scaled to $4\sin^2 \left(\pi p/2N_{\ell}\right)$.
$N_{\ell} = 30$ is the chain length. The data is shown for poly(methyl methacrylate). The line is a fit based using Eq.~\ref{eq:alpha}.
$\alpha_p$ plateaus for $p \ge 11$, which is the typical persistence length of a chain $\ell_{p} \simeq N_{\ell}/p = 2.7$ monomers (or 1.4 nm for PMMA).
The data is taken with permission from Ref.~\cite{MM21mac}. %{\bf change equation number}
}
\label{fig:alphaPMMA}
\end{figure}

Starting from a homogeneous sample consisting of linear polymers, CETM considers the diffusion of energy along a 
chain contour, i.e., between the bonded monomers. This involves multiple hops along a chain
before infrequent energy transfers to the neighboring non--bonded monomer belonging to another chain.
Note that there may also be non--bonded contacts between the two monomers belonging to the same chain, but topologically 
far from one another. However, this will require loop--like conformations of a chain in a dense polymeric system. 
Furthermore, the free energy difference to form such a loop of segment length $\mathcal N$ is given 
by ${\mathcal F}\left(\mathcal N\right) = mk_{\rm B}T \ln\left(\mathcal N\right)$ with a critical exponent $m = 1.95$~\cite{DesBook}.
Within this picture, a loop can only form when it overcomes a free energy barrier of several $k_{\rm B}T$,
which has a very low probability in a dense system. Note also that the CETM method does not distinguish between the
intra-- and intre--molecular non--bonded hopping.

Considering the first and the second neighboring bonded monomer transfers along a chain contour,
The rate of change in the internal energy $\mathcal E$ for any monomer $i$ can then be simply written as~\cite{MM21acsn,MM21mac},
\begin{align}
\label{eq:nnT}
    \frac {{\rm d} {\mathcal E}_i} {{\rm d} t} = c_{\rm m} \frac{{\rm d} T_i}{{\rm d} t}
	%=\frac{\mathrm{d} e_i}{\mathrm{d} t}
    &=G_{\mathrm{b}}(T_{i+1}-2T_i+T_{i-1})\\\nonumber
    &\quad+\tilde G_{\mathrm{b}}(T_{i+2}-4T_{i+1}+6T_i-4T_{i-1}+T_{i-2})\\\nonumber
    &\quad+nG_{\mathrm{nb}}(T_{\mathrm{bulk}}-T_i)\,,
\end{align}
with $G_{\rm b}/c_{\rm m}$, ${\tilde G}_{\rm b}/c_{\rm m}$, and $G_{\rm nb}/c_{\rm m}$ are the bonded, next nearest bonded, and non--bonded energy transfer 
rates, respectively. Here, individually $G$ values are thermal conductances, $c_{\rm m}$ is the specific heat of one monomer, $n$ is the number of non--bonded neighbors, $T_{i}$ is the temperature of the
$i^{\rm th}$ monomer, and $T_{\rm bulk} = 300$ K. 
Following the treatment presented in Ref.~\cite{MM21mac}, diagonalizing Eq.~\ref{eq:nnT} along the chain 
contour will lead to an exponential relaxation of the eigen--modes,
\begin{equation}
\label{eq:expoalp}
	{\hat T}_p\left(t\right) \propto {e}^{-\alpha_p t},
\end{equation}	
with,
\begin{equation}
\label{eq:cos}
	{\hat T}_p (t) = \sum_{i = 0}^{N-1} \{T_i (t) -T_{\rm bulk}\} \cos \left[\frac {p\pi}{N}\left(i+ \frac {1}{2}\right)\right],
\end{equation}
and
\begin{equation}
	\label{eq:alpha}	
	\alpha_p = 4 \frac {G_{\rm b}}{c_{\rm m}}\sin^2\left(\frac {p\pi}{2N}\right) - 16 \frac {{\tilde G}_{\rm b}}{c_{\rm m}}\sin^4\left(\frac {p\pi}{2N}\right)
	+ n \frac {G_{\rm nb}}{c_{\rm m}}.
\end{equation}
In a nutshell, $p$ gives the effective length scale in a system, i.e., a particular $p$ mode corresponds to a length scale of $N_{\ell}/p$.
Fig.~\ref{fig:alphaPMMA} shows the variation in $\alpha_p$ for PMMA.
Fitting the simulation data (symbols) with Eq.~\ref{eq:alpha} gives $G_{\rm b}/G_{\rm nb} \simeq 63$. 
Note also that $G_{\rm b}/G_{\rm nb} \simeq 155$ for a polyethylene (PE) chain~\cite{MM21mac}. This difference is because of the rather bulky 
side group in PMMA that act as an additional scattering center for the energy transfer. This aspect will be discussed at a later stage. 

It can also be appreciated in Fig.~\ref{fig:alphaPMMA} that $\alpha_{p}$ almost plateaus for $4\sin^2\left(\pi p/2N_{\ell}\right) \ge 0.9$ (or $p \ge 11$).
This length scale is comparable to $\ell_{p} \simeq N_{\ell}/p = 2.7$ monomers (or 0.7 nm for PMMA), i.e., a length scale where 
ballistic energy transfer may dominate that is not considered within the formalism of CETM.

The energy transfer rates obtained using Eq.~\ref{eq:alpha} can also be used to get a theoretical estimate of thermal transport coefficient 
within the Heuristic Random--Walk model~\cite{MM21mac},
\begin{equation}\label{eq:kappa_heu}
    \kappa_{\rm HRW} =\frac{\rho_{\rm N}}{6}\left[n G_{\mathrm{nb}}r_{\mathrm{nb}}^2+\left(G_{\mathrm{b}}-4\tilde G_{\mathrm{b}}\right) 
    r_{\mathrm{b}}^2 +\tilde G_{\mathrm{b}} \tilde r_{\mathrm{b}}^2\right].
\end{equation}
Here, $r_{\rm nb}$, $r_{\rm b}$, and ${\tilde r}_{\rm b}$ are the average distances between a monomer 
and its first bonded, second bonded and non-bonded first shell neighboring monomers, respectively. 
It should, however, be noted that $\kappa_{\rm HRW}$ is underestimated for all investigated commodity
polymers~\cite{MM21mac,wu22cms}.

One of the central assumptions in Eq.~\ref{eq:nnT} is that the monomers surrounding the reference chain is kept at a constant 
$T_{\rm bulk} = 300$ K~\cite{MM21mac}. This is certainly a good approximation for the common amorphous polymers,
where the heat leakage between the non--bonded monomers is rather weak and mostly restricted upto the
first non--bonded neighbor. Moreover, in the polymers and lubricants 
under high pressure~\cite{Cahill11PRB,kappaOil,Martin05Sc}, in the confined hydrocarbons~\cite{confinedOil,GAO20Hydro}, 
and/or in the systems where $\pi-\pi$ stacking is dominant~\cite{chen18ScAdv,Smith16AAMI,Shi17AFM}, 
heat leakage between the non--bonded monomers can be significantly enhanced. In these cases, the formalism within CETM may not be directly applicable 
without properly accounting for $T_{\rm bulk}$ that will have a gradient as a function of the radial distance from the central chain.

% \begin{figure*}[ptb]
% \includegraphics[width=0.82\textwidth]{EPSFiles/schematicFiber.pdf}
% \caption{A schematic showing the amorphous configuration (part a) where the non--bonded interactions dominate heat flow.
% In part (b) a schematic of a chain oriented system is shown, where heat flow is dictated by the bonded interactions.}
% \label{fig:schemFiber}
% \end{figure*}

\section{Thermal conductivity of cross--linked polymer networks}

\begin{figure}[ptb]
\includegraphics[width=0.43\textwidth]{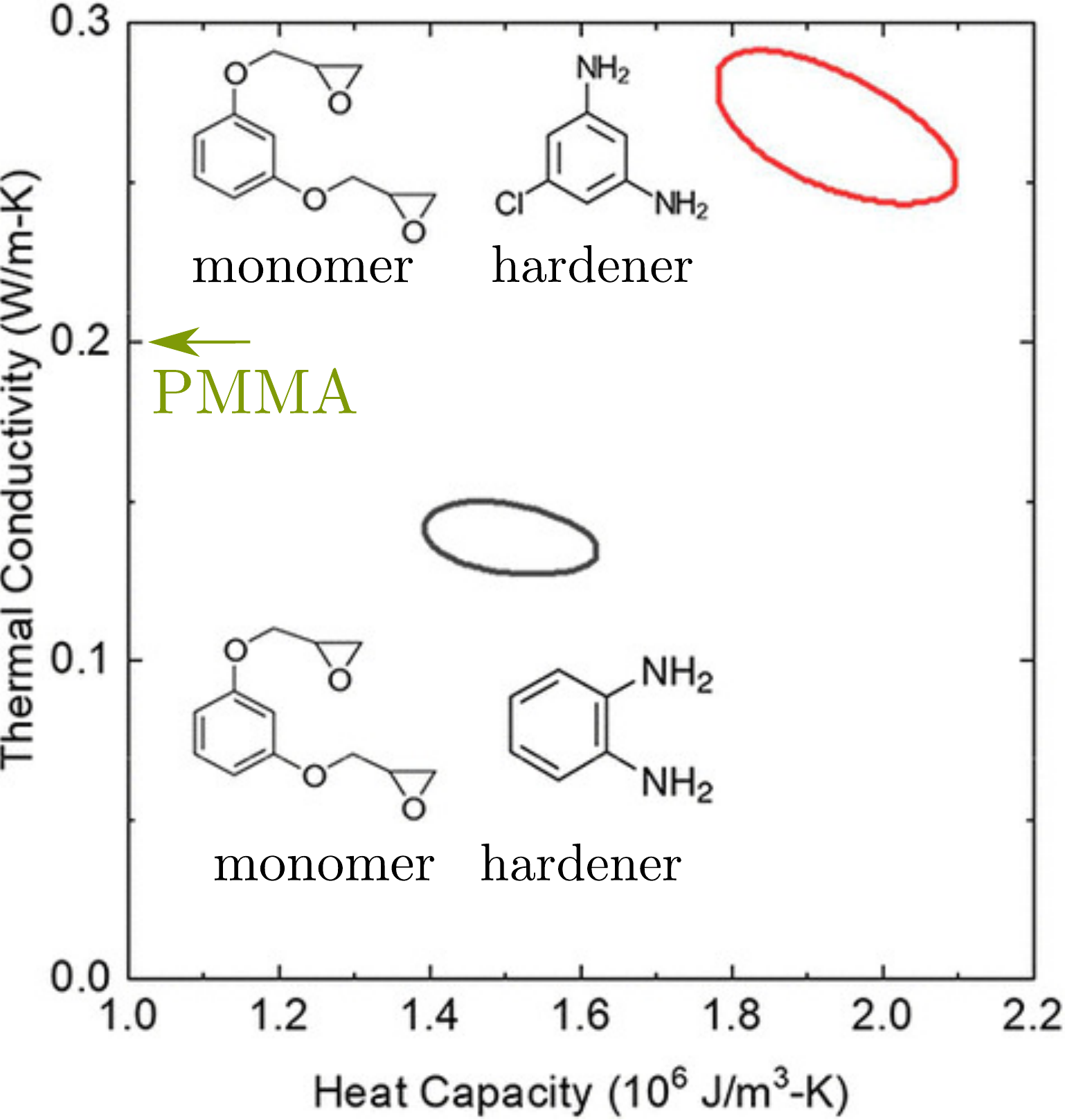}
\caption{Thermal conductivity coefficient $\kappa$ of amine cured epoxy networks with two different hardeners. The arrow 
points at $\kappa \simeq 0.20$ W/Km for a linear poly(methyl methacrylate) system~\cite{cahill16Mac}. An increase of about 1.35
is observed (see the red circle) in comparison to the PMMA data. This figure is reproduced with permission from Ref.~\cite{cahill21acsappm}.}
\label{fig:amineExp}
\end{figure}

A predictive tuning of $\kappa$ purely based on the non--bonded interactions is certainly a non--trivial task, 
if not impossible. Therefore, a more plausible protocol might be to make use of the distinct microscopic interactions (i.e., bonded vs non--bonded),
chain conformation, and possibly also their morphology to understand their effects on $\kappa$. 
In this context, one of the most common classes of polymeric materials where the bonded interactions dominate their properties is the epoxies, 
commonly also referred to as the highly cross--linked polymer (HCP) networks.

In a typical HCP, an individual monomer can form more than two bonds (unlike in a linear chain) 
and thus forms a 3--dimensional bonded network. The HCPs are usually light weight high performance materials with extraordinary mechanical 
response~\cite{White2001,stevens01mac,chenScience,mukherji09pre,Sharifi14JMCA,sirk16SM}, attaining $E$ values that can be 2--3 orders of magnitude 
larger than the common amorphous polymers, consisting of linear chains, and
may provide a suitable materials platform toward the enhancement of $\kappa$~\cite{lv2021effect,DMPRM21,cahill21acsappm,mukherji22acsml}. 
Therefore, recent interest has been devoted in investigating the $\kappa$ behavior in HCPs. 

\subsection{Amine cured epoxies}

One common example of amorphous HCP is the amine cured epoxy networks~\cite{cahill21acsappm,Sharifi14JMCA}, 
where monomers are cross--linked with different amine hardeners with varying stiffness and cross--linker bond length $\ell_{\rm cb}$, 
see the insets in Fig.~\ref{fig:amineExp}. It can be seen from the experimental data in Fig.~\ref{fig:amineExp} that just by changing the hardener, 
$\kappa$ can be tuned by about a factor of two~\cite{cahill21acsappm}. Moreover, even in the best case (shown by the red circles in Fig.~\ref{fig:amineExp}), 
$\kappa$ only increases by a factor of 1.35 in comparison to a linear PMMA, i.e., $\kappa \simeq 0.20$ W/Km~\cite{cahill16Mac,pmmalocalized,pmma14exp}.
The specific enhancement in $\kappa$ is rather small considering that $\kappa$ in epoxies is expected to be dominated by the bonded interactions. 
What causes such a small variation in $\kappa$ for HCPs? To answer this question, direct information about the network micro--structure is 
needed where several competing effects control their physical properties. In this context, obtaining any reasonable information regarding such 
microscopic details is a rather difficult task within the commonly employed experimental techniques. 
Therefore, simulations may be of particular interest in studying the $\kappa$ behavior in epoxy networks, where 
a direct access to the  microscopic network details are reasonable available~\cite{stevens01mac,mukherji09pre,CFA19}. 

Give the chemical specificity of epoxies, one may expect that the all--atom simulations might be the best possible choice.
Moreover, creating a network structure at the all--atom level is difficult and also is computationally expensive, especially when dealing 
with a broad range of system parameters. Complexities get even more elevated because of the large system sizes coupled with 
spacial and temporal heterogeneity~\cite{CFA19}. Therefore, an alternative (and possibly a better choice) is a 
bead--spring type generic simulation technique~\cite{stevens01mac,DMPRM21}. Broadly, generic simulations address the common 
polymer properties that are independent of any specific chemical details and thus a large number of systems can be 
explained within one physical framework~\cite{kremer1990dynamics}. Additionally, tuning the system parameters is rather straightforward within a generic setup,
such as the relative bond lengths, their stiffness, and/or bond orientations that are usually inspired by the underlying 
chemical specific systems~\cite{cahill21acsappm,CamKK01JCP,TakahiroJCP2017}. 

\begin{figure}[ptb]
\includegraphics[width=0.49\textwidth]{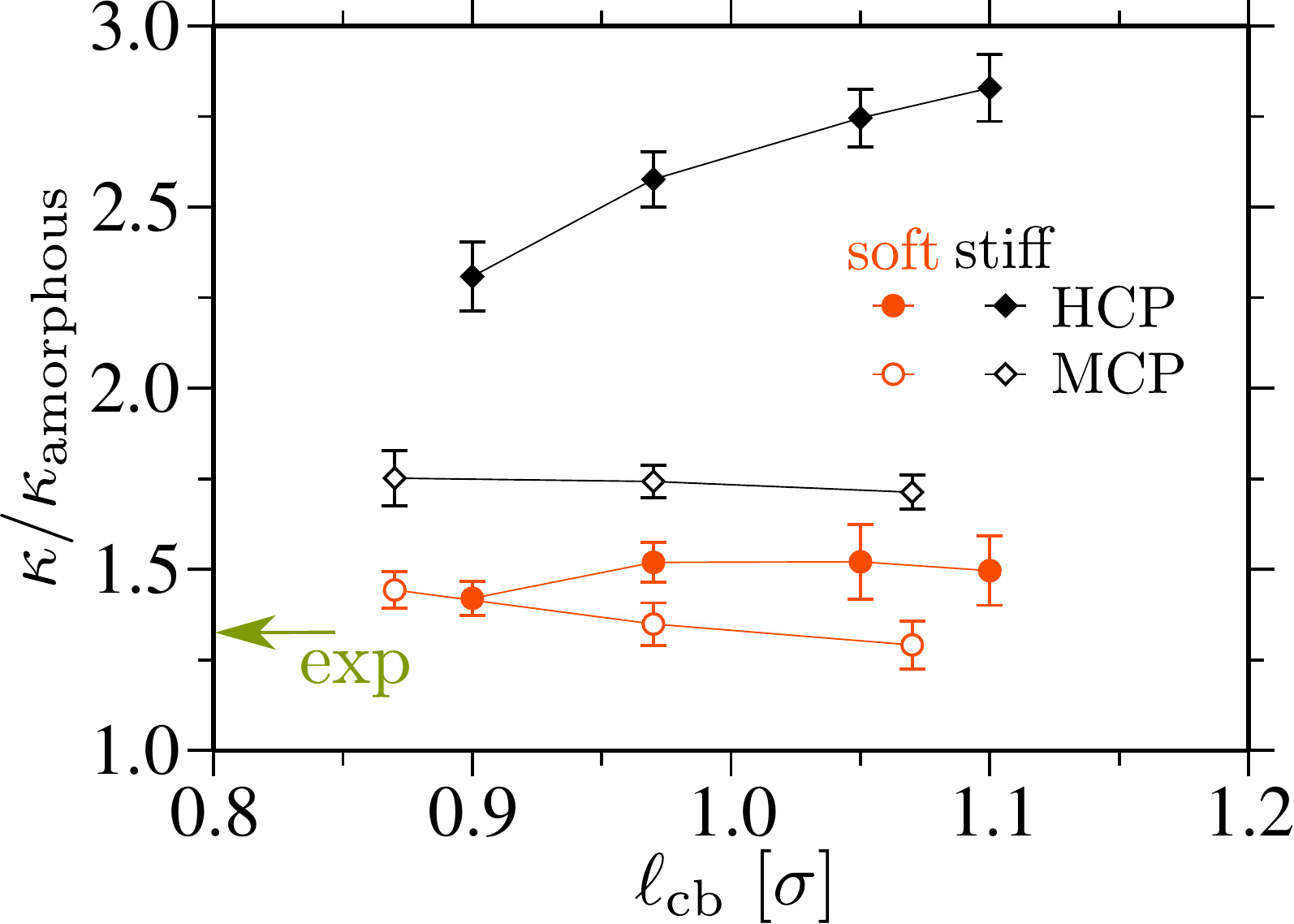}
\caption{Normalized thermal transport coefficient $\kappa$ as a function of the cross-linker (hardener) bond length $\ell_{\rm cb}$.
$\kappa$ is normalized by a linear amorphous polymer $\kappa_{\rm amorphous}$.
Systems are investigated with two different bond stiffnesses, namely; a soft bond representing the amine hardener (orange circles) 
and a stiff bond motivated by the covalent bonds (black diamonds). The data is also shown for two different 
network functionalities. When each monomer can form at most three bonds, a system is termed as the moderate cross--linking (MCP) (solid symbols)
and a system is highly cross--linked (HCP) for the four functional monomers. The arrow 
points at the experimentally observed increase in $\kappa$~\cite{cahill21acsappm}.
The data is taken with permission from Ref.~\cite{DMPRM21}.}
\label{fig:epoxySim}
\end{figure}

\begin{figure}[ptb]
\includegraphics[width=0.43\textwidth]{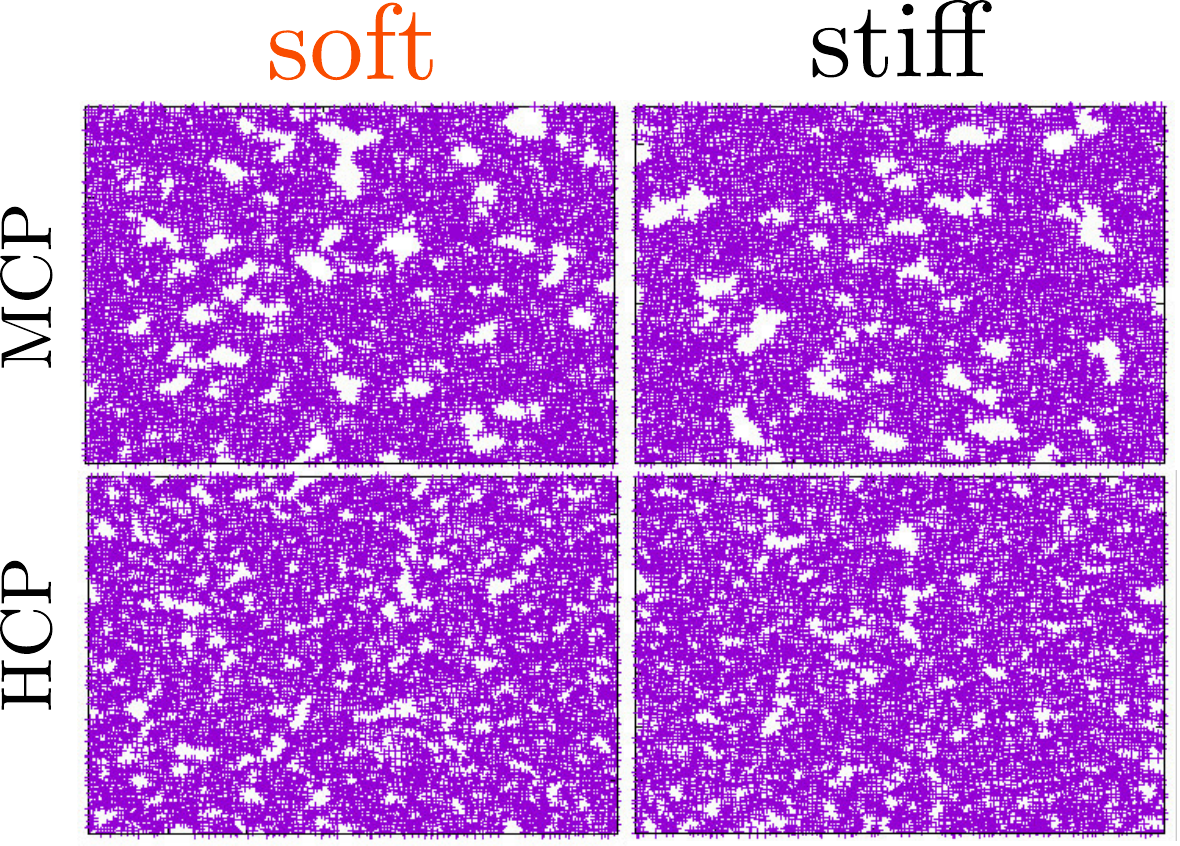}
\caption{Simulation snapshots of two monomer size $\sigma$ thick layer of the samples investigated in Fig.~\ref{fig:epoxySim} and for
a cross-linker (hardener) bond length of $\ell_{\rm cb} = 0.9\sigma$. These figures are
reproduced with permission from Ref.~\cite{DMPRM21}.}
\label{fig:networkMS}
\end{figure}

Fig.~\ref{fig:epoxySim} shows $\kappa$ as a function of $\ell_{\rm cb}$ for a set of model HCPs with different bond stiffness~\cite{DMPRM21}. 
It can be appreciated that the (relatively) soft bonds (representing the amine hardeners) give reasonably consistent values as 
in the experiments~\cite{cahill21acsappm}, i.e., $\kappa/\kappa_{\rm amorphous} \simeq 1.25-1.50$ and it is about 1.35 in Ref.~\cite{cahill21acsappm}
(represented by an arrow in Fig.~\ref{fig:epoxySim}). 

For the stiff bonds, a significant increase in $\kappa$ is observed (represented by the diamond symbols in Fig.~\ref{fig:epoxySim}). 
This large enhancement can be understood by looking into the networks micro--structures. 
From the simulation snapshots in Fig.~\ref{fig:networkMS}, it can be appreciated that there are large voids (or free volume) 
in all the cured samples. Such voids exist when the neighboring monomers form all their bonds pointing out of each other~\cite{mukherji09pre,DMPRM21}
and the monomers along the periphery of a void only interact via vdW forces. These are usually the weak spots within a network,
hence resist the heat flow. %\\\\ Furthermore, t

The observed void sizes are larger for the tri--functional moderately cross--linked polymers (MCP),
while the tetra--functional HCP have relatively smaller free volume. 
A direct consequence is that the MCPs (open symbols) usually have lower $\kappa$ than the HCPs (solid symbols) 
in Fig.~\ref{fig:epoxySim}. In summary, it is not only that the increasing bonded contacts can by default increase $\kappa$. 
Instead the cross--linked bond stiffness, $\ell_{\rm cb}$, and their effects on the network micro--structures control $\kappa$.

\begin{figure}[ptb]
\includegraphics[width=0.4\textwidth]{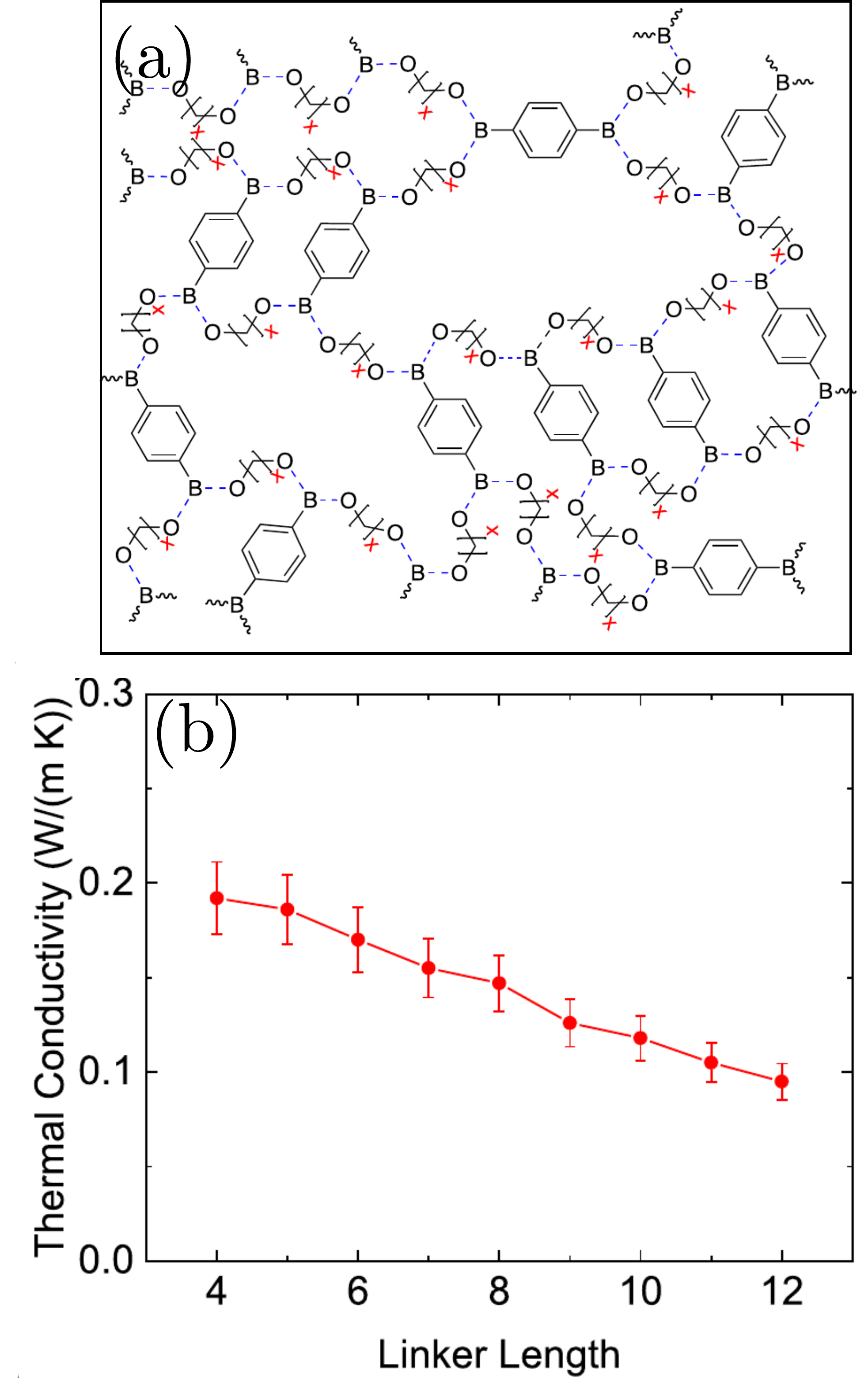}
\caption{Part (a) shows a schematic of network structure with varying ethylene linker length. The corresponding 
thermal conductivity coefficient is shown in part (b). These figures are reproduced with permission from Ref.~\cite{lv2021effect}.}
\label{fig:alkaneHCP}
\end{figure}

\subsection{Ethylene cured epoxies}

Another class is the ethylene cured epoxy networks~\cite{lv2021effect}, where the free volume can be tuned by changing the length
of the ethylene linkers, see the schematic in Fig.~\ref{fig:alkaneHCP}(a). Here, it is important to mention 
that $\ell_{\rm p} \simeq 0.65$ nm for of a PE chain (or equivalent of one ethylene monomer)~\cite{MM21mac}.
When an ethylene linker is longer than $N_{\ell} \ge 4$, it is soft because of its small flexural stiffness. 
The longer the $N_{\ell}$, more flexible is the linker and thus there are also larger free volume in a sample. 
A direct consequence of such a linker is that-- together the free volume and soft linker-- they significantly reduce 
stiffness of a materials and as a result $\kappa$ decreases with increasing $N_{\ell}$, see Fig.~\ref{fig:alkaneHCP}(b).

\section{Thermal conductivity of chain oriented systems}

\subsection{Extended chain configurations}

A typical representing system where bonded interactions dominate $\kappa$ is the polymer fibers~\cite{shen2010polyethylene,luo13acsnano}, 
where individual chain are extended along the direction of heat flow and thus $\kappa$ is dominated 
by the energy transfer between the bonded monomers~\cite{shen2010polyethylene,luo13acsnano,Singh2014,chen19jap,SingleExtend,PEDOT18prm,mukherji24lang}. 
In this context, it has been experimentally reported that a PE fiber can attain $\kappa > 100$ W/Km~\cite{shen2010polyethylene},
which is significantly higher than $\kappa \simeq 0.3$ W/Km for an amorphous PE~\cite{PH}. 

\begin{figure}[h!]
\includegraphics[width=0.46\textwidth]{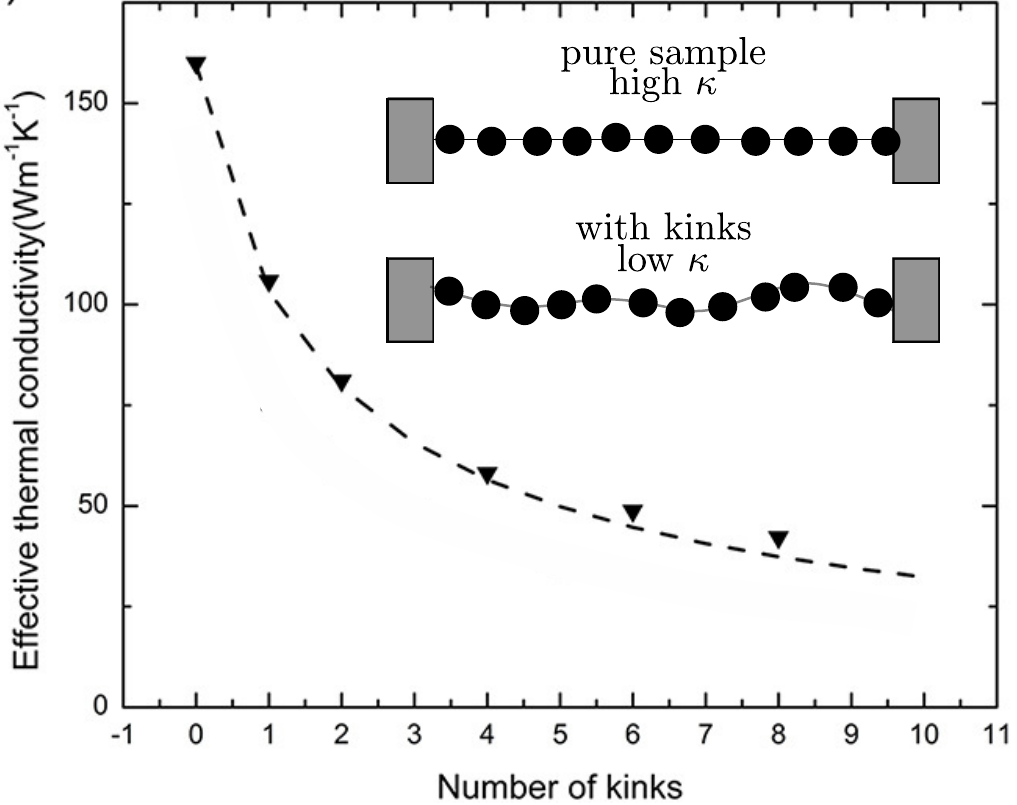}
\caption{The main panel shows the thermal transport coefficient $\kappa$ with the number of kinks along a polyethylene (PE) chain. 
The main figure is reproduced with permission from Ref.~\cite{chen19jap}.
Insets show two schematics representing of a pure sample without kinks and a same sample with a few kinks. }
\label{fig:k_kinks}
\end{figure}

A closer look at an extended chain configuration reveals that it can be viewed as a quasi one--dimensional (Q1--D) 
crystalline material~\cite{nanotube2016}. This is a direct consequence of the periodic arrangement 
of monomers along a chain backbone, see the top schematic in the inset of Fig.~\ref{fig:k_kinks}.
In such a system, phonons carry a heat current and the coupling strength between the lattice sites is dictated by the bonded interactions. 
Usually a pure (pristine) sample has a large $\Lambda$ and thus also a high $\kappa \simeq 160$ W/Km~\cite{chen19jap}, see the main panel in Fig.~\ref{fig:k_kinks}. 
However, whenever there appears a kink or a bend along a chain contour (see the bottom schematic in the inset of Fig.~\ref{fig:k_kinks}), 
it scatters the phonons. The larger the number of kinks along a chain, the larger the resistance for heat flow and thus a lower $\kappa$.
This picture is well supported by the simulation results of an extended PE chain with varying number of kinks, 
see Fig.~\ref{fig:k_kinks}.

\begin{figure}[ptb]
\includegraphics[width=0.46\textwidth]{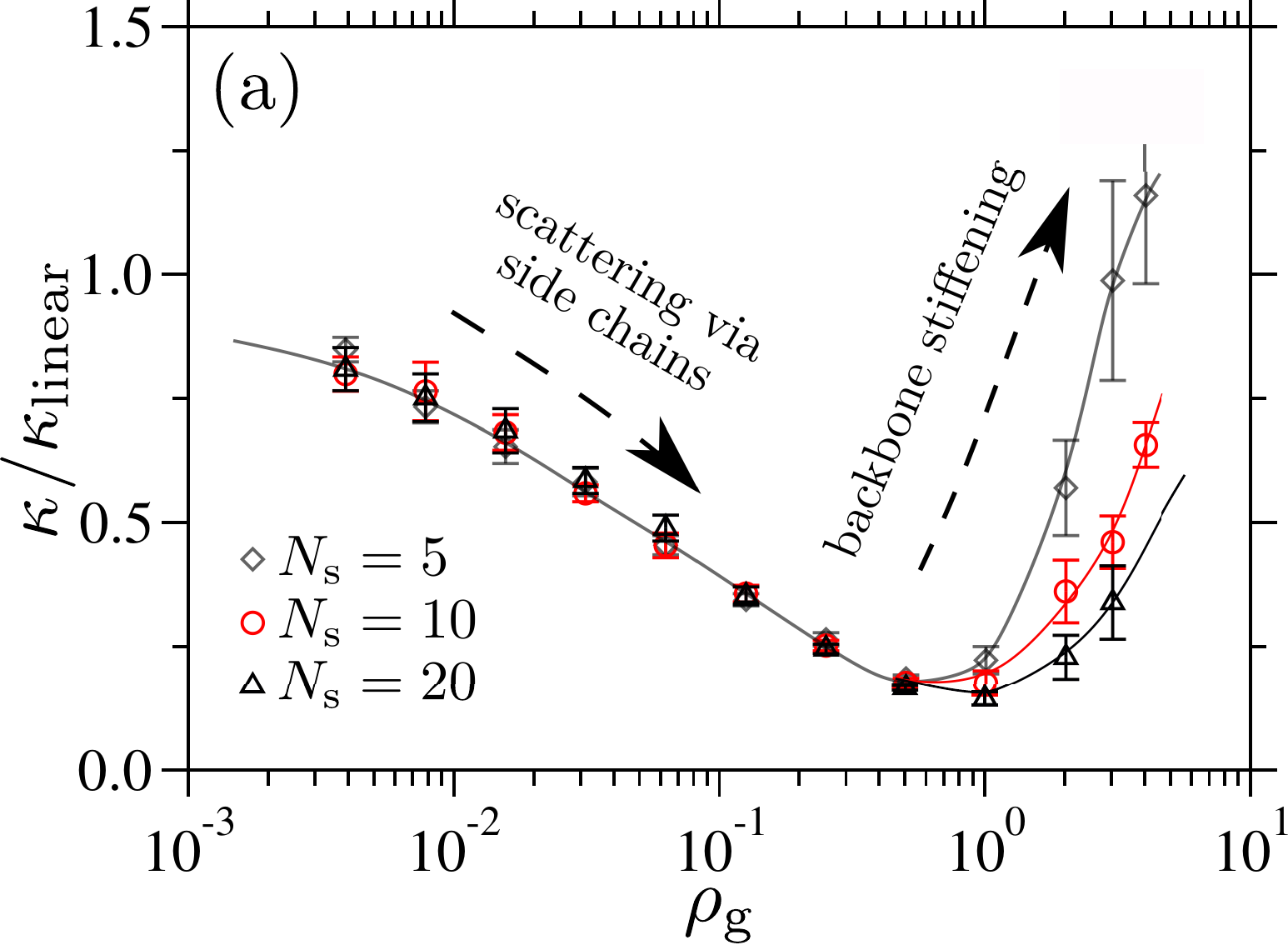}
\includegraphics[width=0.46\textwidth]{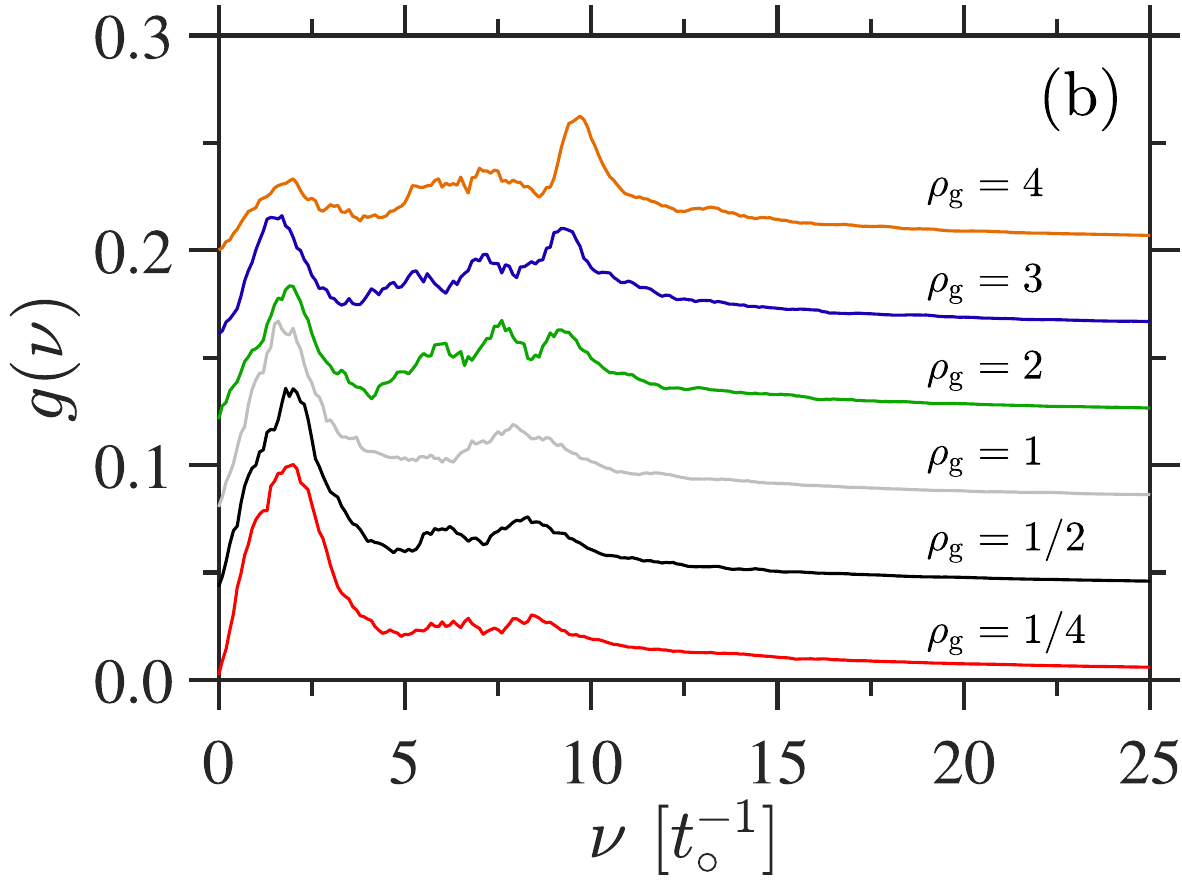}
\caption{Part (a) shows the normalized thermal transport coefficient $\kappa$ as a function of the side chain grafting density $\rho_{\rm g}$
of the bottle--brush polymers (BBP). The data is shown for three different side chain lengths $N_{\rm s}$.
The data is normalized with $\kappa_{\rm linear}$ of a corresponding linear chain, i.e., at $\rho_{\rm g} = 0$.
Part (b) shows the representative vibrational density of states $g(\nu)$ for the backbone of the BBPs 
for six different $\rho_{\rm g}$. These figures are reproduced with permission from Ref.~\cite{mukherji24lang}.}
\label{fig:BBkappa}
\end{figure}

To investigate the effects of kinks on $\kappa$, different PE configurations were specifically engineered~\cite{chen19jap}. 
However, a natural system where the number of kinks and the backbone stiffness can be controlled almost \textit{at will} is the bottle--brush polymers (BBP)~\cite{marquesThesis1989,binderJCP2011}. 
A polymer is referred to as a BBP when a linear polymer of length $N_{\ell}$ is grafted with the side chains with varying
length $N_{\rm s}$ and grafting density $\rho_{\rm g}$. Here, $\rho_{\rm g}$ is defined as the 
number of side chains grafted per backbone monomer. For example, if every backbone monomer is grafted with one side chain, then $\rho_{\rm g} = 1$.
BBPs are of interest because of their potential in designing one--dimensional organic nano--crystals~\cite{nanotube2016}.

In a BBP, the backbone flexural stiffness (controlling the number of kinks along a chain) is dictated by $N_{\rm s}$ and $\rho_{\rm g}$~\cite{marquesThesis1989,binderJCP2011}.
The heat management in the BBPs is of particular interest because their $\kappa$ is dictated by two competing effects~\cite{kappaHaoMaa,mukherji24lang}: 
(1) The presence of side chains act as the pathways for heat leakage that effectively reduce $\kappa$. 
(2) The side chains increase the backbone flexural stiffness and thus there exists less kinks (or defects) along the backbone, 
which effectively increases $\kappa$. To investigate the extent by which these two effects control $\kappa$, recent simulations
have been performed using a generic model. The representative data is shown in Fig.~\ref{fig:BBkappa}(a).

It can be appreciated in Fig.~\ref{fig:BBkappa}(a) that $\kappa$ shows a non--monotonic variation with $\rho_{\rm g}$, where
two regimes are clearly visible: For $\rho_{\rm g} \le 1$, scattering because of the side chains reduces $\kappa$, while the 
backbone stiffening via side chains increases $\kappa$ for $\rho_{\rm g} > 1$~\cite{mukherji24lang}. This backbone stiffening scenario is also 
supported by $g(\nu)$. It can be seen in Fig.~\ref{fig:BBkappa}(b) that there are almost indistinguishable changes in $g(\nu)$ 
for $\rho_{\rm g} \le 1$. Moreover, when $\rho_{\rm g} > 1$ a peak becomes more prominent around $\nu \simeq 9-10 t_{\circ}^{-1}$. 
This is associated with the flexural stiffness that increases with increasing $\rho_{\rm g}$, 
as revealed by the shift in this peak towards the higher $\nu$ values. The full--width--of--half--maxima $\nu_{\rm FWHM}$ also 
decreases with increasing $\rho_{\rm g}$ and thus increases the phonon life time $\tau \propto 1/\nu_{\rm FWHM}$ (or $\kappa$).

\subsection{Molecular forests}

The knock down in $\kappa$ via kinks is a concept that can also be helpful in dictating the heat flow in more complex 
molecular assemblies. One example is molecular forest, where Q1--D are grafted perpendicularly on a
surface forming a two dimensional assembly, such as the forests of CNT~\cite{CNTForest,HeatTrap,CNTBundles}, silicon nanowires~\cite{CahillSiForest,SiForest18Nanotech}, 
and/or polymers~\cite{shen2010polyethylene,PEDOT18prm,bhardwaj2021thermal}.
These forests often exhibit intriguing and counter--intuitive physical behavior. 
In this context, it had been experimentally reported that-- while a single CNT has $\kappa \ge 10^3$ W/Km~\cite{CNT1,CNT2,CNT3,CNT4}, 
the same CNT in a forest shows a drastic reduction in $\kappa$~\cite{CNTForest,HeatTrap,CNTBundles}. %by 3--4 orders of magnitude.
This phenomenon is commonly referred to as the {\it heat trap} effect (HTE) in the carbon nanotube (CNT) forests~\cite{HeatTrap}.

\begin{figure}[ptb]
\includegraphics[width=0.49\textwidth]{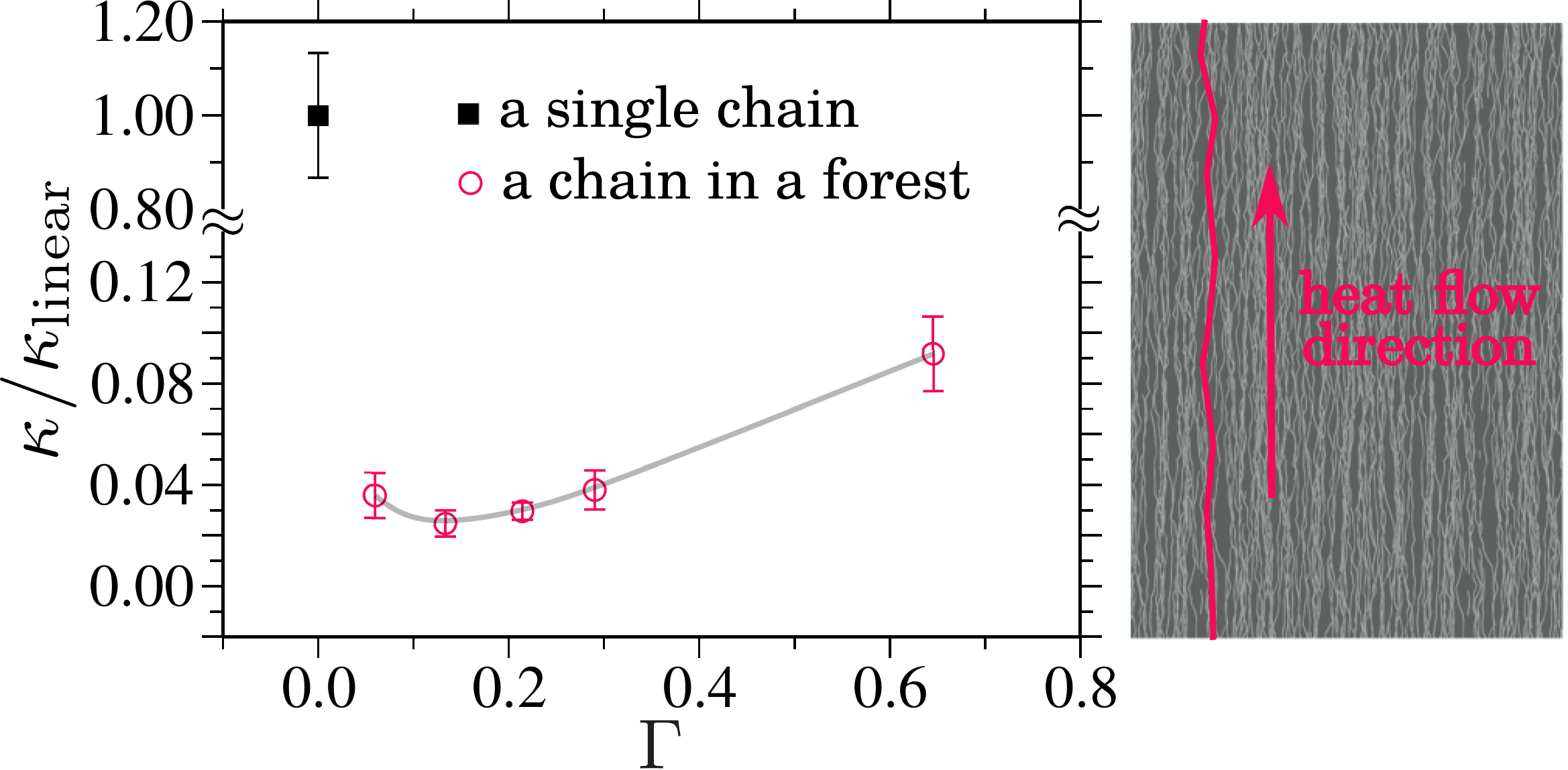}
\caption{The left panel shows the normalized thermal transport coefficient $\kappa$ of a chain in a forest as a function of the grafting 
density of the chains $\Gamma$. The data is normalized with $\kappa_{\rm linear}$ of an isolated linear chain.
The right panel shows a representative simulation highlighting the direction of heat flow.
These figures are adapted with permission from Ref.~\cite{bhardwaj2021thermal}.}
\label{fig:heattrap}
\end{figure}

Even when the counter--intuitive HTE phenomenon was known for over a decade, there existed no clear understanding of this behavior.
Here, the simple concepts known from soft matter physics turned out to be reasonably useful in understanding
certain aspects of a hard matter problem of complex molecular assemblies. For this purpose, generic simulations were performed~\cite{bhardwaj2021thermal}.
The key assumption in this model is that a Q1--D is considered as a single extended polymer chain and thus
represents a molecular forest as a polymer brush. The only input parameter in such a modelling approach is $\ell_{\rm p}$.
In this context, it was readily observed that a CNT can be characterized by their bending stiffness, as measured
in terms of $\ell_{\rm p}$, that increases with CNT diameter $d$~\cite{CNTLP}. For example, a CNT of $d = 1$ nm has $\ell_{\rm p} = 50-60~\mu$m. 
Within this picture, a CNT forest of $2$ mm height can have about $30-40\ell_{\rm p}$
or as many number of kinks. Note that in this simple argument we do not discuss the effect of grafting density $\Gamma$. 
The representative data and the corresponding simulation snapshot is shown in Fig.~\ref{fig:heattrap}.
As expected, a chain in a forest shows a significant reduction in $\kappa$~\cite{bhardwaj2021thermal}, see the left panel in Fig.~\ref{fig:heattrap}. 
As discussed above, this knock down is direct consequence of the kinks that act as the scattering centers for the heat flow.
The kinks are also evident from the simulation snapshot in the right panel of Fig.~\ref{fig:heattrap}.

\section{Thermal conductivity of crystalline polymers}

A somewhat different class to the amorphous polymers is the polymers with certain degree of crystalline order, 
where the long range order facilitates phonon propagation that carry a heat current and thus results in 
an enhanced rate of energy transfer. The typical examples include liquid--crystalline 
materials~\cite{HA2013113,CisTrans19PNAS,ishida22NL}, poly--peptide sequences~\cite{tomko2018tunable}, 
and/or semi--crystalline polymers~\cite{chen18ScAdv,CrystalPol19,luo13acsnano}. 

\subsection{Liquid crystalline polymers}

Liquid crystals usually have $\kappa \simeq 0.3$ W/Km~\cite{CisTrans19PNAS,ishida22NL}, i.e., similar to the commodity
polymers. However, one of the advantages of a liquid crystalline material, such as the azobenzene--based liquid crystals, 
is that an azobenzene undergoes a re--entrant {\it trans}--to--{\it cis} transition when they are exposed to near--ultraviolet light~\cite{ctUltra1,ctultra2}. 
Such a transition also alters the molecular order in a liquid crystal and hence $\kappa$ switches 
between 0.1 W/Km (in {\it cis} state) and 0.3 W/Km (in {\it trans} state)~\cite{CisTrans19PNAS}.
This can simply be viewed as a light responsive thermal switch.

When a liquid crystalline polymer is cross--linked with the ethylene linkers, they give very interesting 
and counter--intuitive trends in $\kappa$~\cite{Koda2013,cahill22pnas}. For example, earlier experimental studies have reported 
that $\kappa$ shows a zig--zag variation with increasing $N_{\ell}$, varying between 1.0--0.2 W/Km~\cite{cahill22pnas}. 
For an even number of carbon atoms in a linker, $\kappa$ always has a higher 
value than the next system with a linker with an odd number of carbon atoms. This behavior is commonly known as the odd--even effect in $\kappa$,
which was initially reported in a simulation study~\cite{Koda2013}. Moreover, a more detailed investigation is recently reported~\cite{cahill22pnas}. 
Such an odd--even effect is also well--known in various other properties of the liquid crystalline polymers~\cite{oddEven75JCP,oddEven19MH}.
While these studies gave very nice insight into the $\kappa$ behavior of these complex systems,
an exact molecular level understanding of such a non--trivial odd--even effect is still somewhat lacking.

\subsection{Conjugated polymers}

Another polymeric system, where crystalline ordering is probably most important, is the conjugated polymer
because they are often used under the high temperature conditions. The crystalline order in such a system
is because of the $\pi-\pi$ stacking of their backbone consisting of the aromatic structures~\cite{nancy22rev,chen18ScAdv,Smith16AAMI,Shi17AFM}. 
In this context, a significantly large value of $\kappa \to 2.0$ W/Km was reported in poly(3-hexylthiophene) (P3HT)~\cite{chen18ScAdv}.
This $\kappa$ can also be further increased by blending with the multi--wall CNTs~\cite{Smith16AAMI}. The latter
study reported a non--monotonic variation with $\phi_{\rm CNT}$, reaching a maximum value of about 5.0 W/Km around $\phi_{\rm CNT} \simeq 30\%$.

\subsection{Poly--peptide sequences}

A natural soft matter that shows structural order is a poly--peptide sequence. 
Previous experimental results have shown that-- by controlling the specific amino acid residues
along a poly--peptide sequence, one can significantly alter its degree of secondary structure $d_{\rm s}$. 
In such a system, an enhancement of up to $\kappa \simeq 1.5$ W/Km is observed in a hydrated poly--peptide~\cite{tomko2018tunable}. 
Note also that $\kappa \simeq 0.6$ W/Km for water and hence the observed increase is a direct consequence of $d_{\rm s}$.
Here, however, it is important to mention that the controlled synthesis (aka precision polymerization) of a specific 
poly--peptide sequence is a grand challenge and they are commercially expensive~\cite{weil20rev}. 

\subsection{Semi--crystalline polymers}

\begin{figure*}[ptb]
\includegraphics[width=0.997\textwidth]{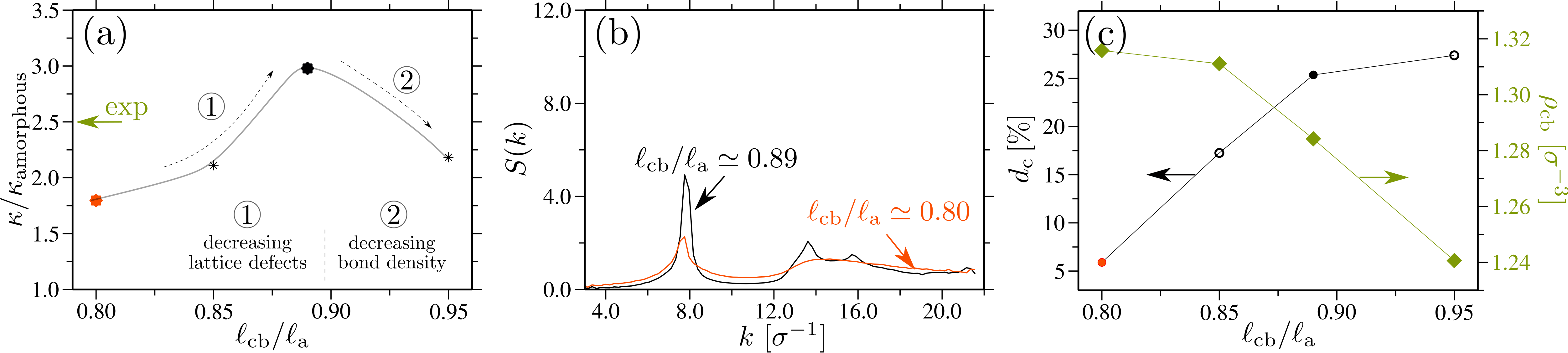}
\caption{Part (a) shows the normalized thermal transport coefficient $\kappa$ of a semi--crystalline network as a 
function of cross--linker bond length $\ell_{\rm cb}$. $\kappa$ is normalized by the 
thermal transport coefficient of a corresponding amorphous sample $\kappa_{\rm amorphous}$.
The corresponding structure factor for two different samples are shown in part (b). In part (c), degree of crystallinity $d_{\rm c}$ and the 
cross--linker bond density $\rho_{\rm cb}$ are shown. These figures are reproduced with permission from Ref.~\cite{mukherji22acsml}.
\label{fig:semicEpoxy}}
\end{figure*}

%Given the discussions above, a 
A more plausible alternative to poly--peptides might be the semi--crystalline synthetic commodity polymers, 
such as the PVA, PLA, and PE systems~\cite{semicrystalline2,semicrystallineRev}. 
While $\kappa$ of a semi--crystalline sample can be rather large because of the long range order, 
it may be inferred that if they are cross--linked, the combination of these two effects might lead to an even greater 
increase in $\kappa$ than the bare amorphous polymers. Motivated by this, a recent experimental study has 
reported that a cross--linked semi--crystalline network can only achieve an increase of up to a factor of 
2.5 times than the pure PMMA sample, i.e., $\kappa \simeq 0.5$ W/Km~\cite{Cahill21SemiCrys}. This rather surprising behavior was also 
investigated in a simulation study, where a similar increase in $\kappa$ was observed
for a critical $\ell_{\rm cb}$, see Fig.~\ref{fig:semicEpoxy}(a). %for a specific $\ell_{\rm cb}$. 

A maxima in $\kappa$ is only observed when $\ell_{\rm cb}$ is comparable to the lattice constant $\ell_{\rm a}$, such that the 
degree of crystallinity $d_{\rm c}$ increases with $\ell_{\rm cb}$, as revealed by the peak heights in the scattering function in Fig.~\ref{fig:semicEpoxy}(b)~\cite{mukherji22acsml}. 
When $\ell_{\rm cb}$ increases beyond a certain threshold (i.e., $\ell_{\rm cb}/\ell_{\rm a} \ge 0.9$)--
in one hand $d_{\rm c}$ increases only slightly, on the other hand there is a large decrease in the 
bond density $\rho_{\rm cb}$, see Fig.~\ref{fig:semicEpoxy}(c). To summarize the data in Fig.~\ref{fig:semicEpoxy}(a), 
the observed initial increase in $\kappa$ for $\ell_{\rm cb}/\ell_{\rm a} \le 0.9$ is due to the increased $d_{\rm c}$ and 
the decrease in $\kappa$ for $\ell_{\rm cb}/\ell_{\rm a} \ge 0.9$ is due to the 
reduced $\rho_{\rm cb}$. This readily suggests that a delicate combination of $d_{\rm crystal}$ and $\rho_{\rm cb}$ controls $\kappa$~\cite{Cahill21SemiCrys,mukherji22acsml}.

\section{Thermal conductivity of polyelectrolytes}

In the preceding sections, a short overview of the $\kappa$ behavior in neutral polymers are presented. 
Possible ideas are also discussed that can be used to tune $\kappa$ in amorphous systems by macromolecular
engineering. However, there are systems where electrostatic interaction also plays an important role, 
examples include but are not limited to, organic (soft) electronics~\cite{Smith16AAMI,chen18ScAdv,Shi17AFM}, bio--inspired materials~\cite{tomko2018tunable}, 
and flexible chips~\cite{Chips3,chips2,chips1}. 
In these systems it is always desirable to attain a large $\kappa$ that can act as a heat sink and thus improves device 
performance/durability. Because of this need, extensive efforts have been devoted in studying the $\kappa$ behavior 
in electrostatically modified polymers~\cite{AShank17ScAdv,Luo19JPCC}.

One of the classical examples of polyelectrolytes is the modified PAA with varying degree of ionization. 
Here, a recent experimental study has investigated the effect of pH on ionized PAA, which reported $\kappa \to 1.2$ W/Km~\cite{AShank17ScAdv}.
This is an enhancement of about 3--4 times than the neutral amorphous PAA, where $\kappa \simeq 0.20-0.37$ W/Km~\cite{cahill16Mac,Pipe15NMat,AShank17ScAdv}. 
This enhancement was also coupled with a significant increase in the materials stiffness $E$, i.e., consistent with the 
predictions of the MTCM that $\kappa \propto E$~\cite{Cahill90PRB,Braun18AM}. The increased $E$ was predominantly because
electrostatic interaction stretches a PAA and thus makes bonded interaction more dominant than in the case 
of an uncharged PAA system. This characteristic extension of an ionized chain is also consistent with the earlier 
studies investigating the effective stretching and $\ell_{\rm p}$ of polyelectrolytes~\cite{Buehler12mac,PolelectMac,MuthukumarRev}.

The $\kappa$ behavior in the polyelectrolytes suggest that the influence of the electrostatic interaction is rather indirect, i.e., 
they help stretch a chain and thus the bonded interactions become more dominant, which increases $\kappa$.
Something may speak in this favor that the electrostatics alone do not influence $\kappa$,
as in the case of the ionic liquid consisting of small molecules~\cite{nancy22mol}, where $\kappa \simeq 0.2$ W/Km~\cite{Tomida18}.

\section{Thermal transport in smart responsive polymers}

The backbone structure of the commodity polymers are commonly dominated by the C--C covalent bonds, see Fig.~\ref{fig:struct}.
Such a bond is extremely strong with its strength of about 80$k_{\rm B}T$ and thus these bonds live forever under the unperturbed 
environmental conditions. This creates severe ecological problems, which get even worse when dealing with water 
insoluble polymers, as shown by a few example in the top panel of Fig.~\ref{fig:struct}. 
This is one of the main reasons why the recent interest has been diverted to water soluble (H--bonded) polymers, 
referred to as the ``smart" polymers~\cite{winnik15rev,Mukherji20AR}, as shown by a few examples in the bottom panel of Fig.~\ref{fig:struct}. 
Additionally, it is also preferred if a polymer can be bio--degradable and/or pH responsive, such as the acetal--linked 
copolymers~\cite{koberstein16mac}. 

A polymer is referred to as a ``smart" responsive when a small change in the external stimuli can significantly alter 
their structure, function, and stability. These stimuli can be temperature~\cite{chi98prl,winnik15rev,tiago18pccp,Mukherji20AR}, 
pressure~\cite{richtering13Press,Mukherji15SM,Papadakis23Lang}, pH~\cite{koberstein16mac}, light~\cite{LS1,LS2,LS3}, and/or cosolvent~\cite{muthukumar91mac,winnik90mac,chi01prl,Mukherji14NC}. 
One common example of smart polymer is PNIPAM that shows a coil--to--globule transition in water around $T_{\ell} \simeq 305$ K (or $32^{\circ}$ C)~\cite{chi98prl,winnik15rev}. 
This is a typical lower critical solution (LCST) behavior~\cite{winnik15rev,koberstein16mac,TakahiroJCP2017}
driven by the solvent entropy~\cite{DoiBook,DGbook,DesBook}.

The fast conformational switching of PNIPAM in water may be extremely useful in the thermal applications.
Thermal switching is one such application that controls heat flow in various systems, including, but are not limited to, 
thermoelectric conversion, energy storage, space technology, and sensing~\cite{chen11NC,pnipamLCSTkappa,tian18acsml,thermS19jpcc}. 
In this context, the conventional thermal switches often suffer from their slow transition rates and thus also have poor performance.
\begin{figure}[ptb]
\includegraphics[width=0.46\textwidth,angle=0]{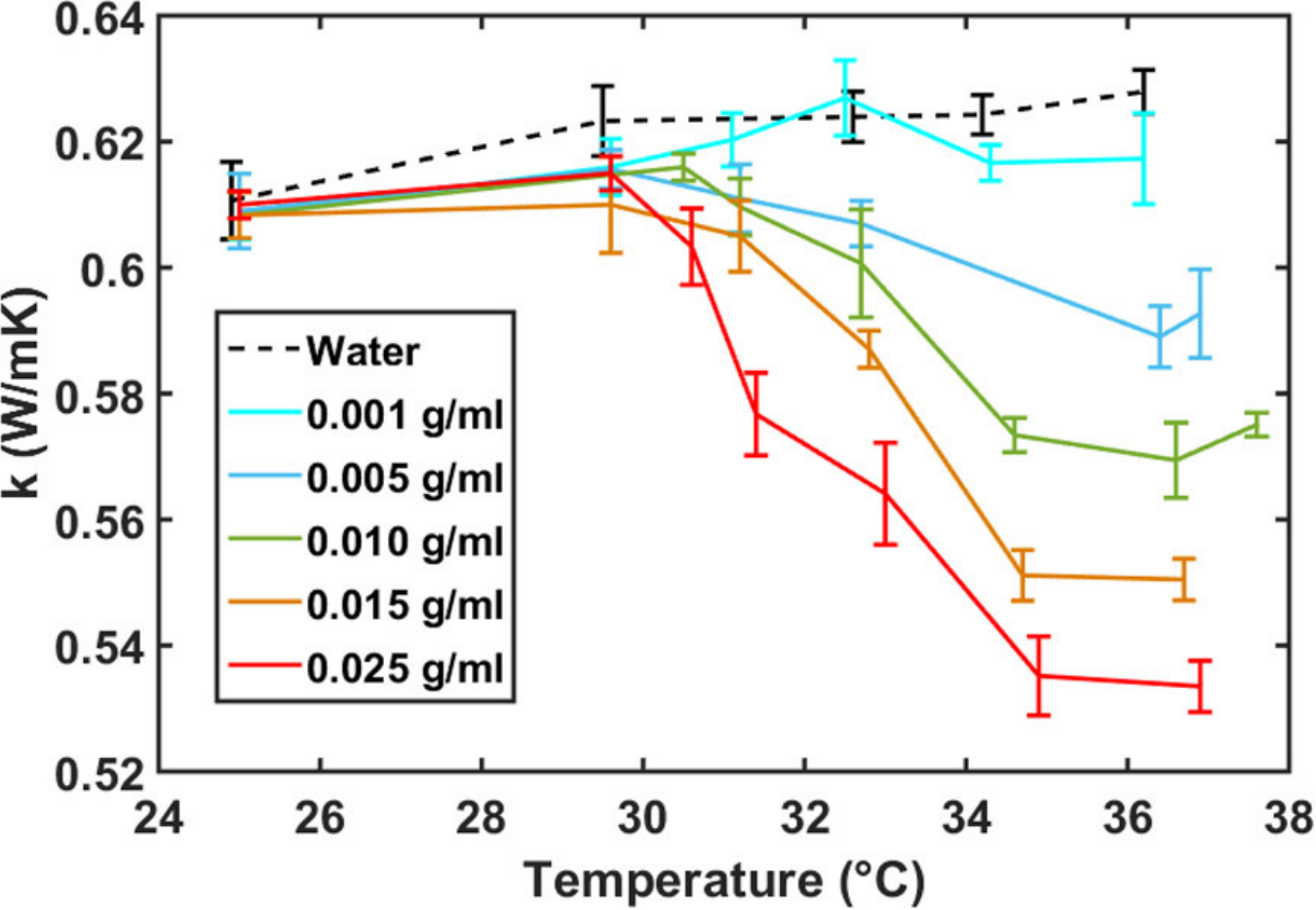}
\caption{The thermal transport coefficient $\kappa$ as a function of temperature $T$ for the 
aqueous poly(N--isopropylacrylamide) (PNIPAM) solutions with changing PNIPAM concentrations. 
This figure is reproduced with permission from the American Chemical Society~\cite{tian18acsml}.
\label{fig:TSw}}
\end{figure}
Recently there has been considerable interest in studying $\kappa$ in the smart polymer with a goal to attain a fast switching 
in $\kappa$~\cite{pnipamLCSTkappa,tian18acsml,thermS19jpcc}. %,thermPAAM}. 
In particular, experimental studies in the aqueous solutions of PNIPAM~\cite{tian18acsml} and PNIPAM--based hydrogels~\cite{thermS19jpcc} 
have shown that their $\kappa$ behavior follow the same trend as the LCST transition around $T_{\ell} \simeq 305$ K 
(or $32^{\circ}$ C)~\cite{chi98prl,winnik15rev}, see Fig.~\ref{fig:TSw}.

It is important to note that $\kappa$ increases with $T$ in the liquids and in the amorphous materials~\cite{Cahill90PRB,wu22cms,MukherjiarXiv,plakappa,plapmmaExp,pmma14exp,pmmalocalized,keb09jap}
because of an increased vibrations. %As a result, the typically macroscopic $\kappa$ increases with increasing $T$~\cite{wu22cms,keb09jap}. 
This behavior is also visible in pure water, where a weak increase in $\kappa$ is observed with $T$, see the black data set in Fig.~\ref{fig:TSw}.
Moreover, the sudden drop in $\kappa$ around $T \ge 303$ K (or $30^{\circ}$ C) for the PNIPAM concentration above $5 \times 10^{-3}$ g/mL
is predominantly due to the coil--to--globule transition of PNIPAM. This drop is likely due to the loss in the number of hydrogen bonds needed 
to stabilize a PNIPAM configuration and the resultant breakage of the water caging around PNIPAM~\cite{tian18acsml,thermS19jpcc}. 
These broken water--PNIPAM H-Bonds effectively creates weak interfaces that act as resistance for the heat flow.
Contrary to these results, another experimental study has reported an opposite trend for the concentrated PNIPAM solutions, 
i.e., $\kappa$ increases above $T_{\rm cloud}$~\cite{pnipamLCSTkappa}. This is simply because a chain under
a high concentration does not collapse into a globule, instead it remains rather expanded surrounded by the other neighboring chains 
and hence the $\kappa$ behavior become dominated by the energy transfer between the bonded monomers.
Furthermore, these distinct results highlight that the polymer concentration, chain size at a given concentration, 
relative interaction/coordination (monomer--monomer, monomer--solvent, and solvent--solvent), water tetrahedrality around 
a PNIPAM and a delicate balance between these effects play key role in dictating $\kappa$ behavior in the polymer solutions. 
A detailed understanding of such effects on the $\kappa$ behavior is a rather open discussion.

Lastly, it might also be important to highlight another (possible) system for the thermal switching application. 
In this context, elastin--like poly--peptides (ELP) are a modern class of biomimetic polymers that also shows LCST transition~\cite{chilkoti99nb,elp2}.
One important aspect of the ELPs is their proline isomerization (ProI) that can have either a {\it cis} or a {\it trans} conformation, which can 
dictate their relative conformations~\cite{CT1,CT2,Mukherji24mrc}. The free energy barrier of such a {\it cis}--to--{\it trans} transition is 
about 30$k_{\rm B}T$, which the free energy difference between these two states in only about 2$k_{\rm B} T$ and thus can be switched via light.
Given the above discussion, ELPs with ProI may also be alternatively used as a thermal switch, similar to that in the liquid crystalline materials~\cite{CisTrans19PNAS,ishida22NL}.

\section{Concluding remarks}

Ever since the seminal publication of Hermann Staudinger~\cite{staudinger1920}, the field of polymer science has traversed a long journey 
with many new interesting developments for the future design of advanced functional materials. In the constant quest to find 
new polymeric materials with improved performance, significant attention has been devoted within the field thermal transport 
of polymers over the last 2--3 decades. Especially because the polymeric plastics usually have very low thermal conductivity coefficient $\kappa$, which is 
typically a few orders of magnitude smaller than the common crystals. Here, one of the grand challenges is to attain a predictive tuning
of $\kappa$ via macromolecular engineering. In the context, experiments have investigated a plethora of systems that
include-- linear polymers, symmetric and asymmetric polymer blends, polymer composites, cross--linked networks, polymer fibers, 
crystalline polymers and electrostatically modified polymers, to name a few. Motivated by these studies, computational 
studies have also been conducted to establish a structure--property relationship in polymers within the context of their $\kappa$ behavior.

While it is certainly rather difficult to address all aspects of a huge field of research within one short overview,
in this work an attempt has been made to highlight some of the latest developments in the field of heat conductivity 
in polymers and polymeric materials. In particular, computational results are discussed within the context of the complementary 
experiments with a goal to establish a detailed microscopic understanding that dictates macroscopic polymer properties. 
Available theoretical models are also discussed that may pave the way to guide the future experimental and/or simulation studies.

Some discussions are also presented that showed that the simple concepts know from the basic polymer (soft matter) science~\cite{bhardwaj2021thermal} can be
used to understand a complex problem from an opposite class of hard matter physics~\cite{HeatTrap,CNTForest}. This further highlights why polymer science
is such a vibrant and active field of research which is not only restricted within the soft matter community. Rather, it reaches 
across a wide range of interdisciplinary fields. \\

\noindent{\bf Acknowledgement:} The contents presented in this review have greatly benefited from the discussions with many colleagues. 
In particular, the development of two key concepts presented here would not have been possible without 
very fruitful collaborations with Martin M\"user and Marcus M\"uller, whom I take this opportunity to gratefully acknowledge.

This draft is a contribution towards a special issue to celebrate the 40th anniversary of Max Planck Institute for Polymer Research. 
I take this opportunity to gratefully acknowledge very fruitful continual collaborations with many MPIP colleagues,
especially Kurt Kremer for numerous stimulating discussions that led to the foundation of my works in MPIP and also after.

I further thank Kyle Monkman for useful comments on this draft.\\

\noindent{\bf Conflict of interest:} The author declares no conflicting financial interest.\\

\noindent{\bf Copyright permission statement:} Copyright permissions are obtained for all figures used in this review.

\bibliographystyle{ieeetr}
\bibliography{reference.bib}

\end{document}